\documentclass[runningheads, 11pt]{article}

\usepackage{times}
\usepackage{physics}
\usepackage{StyleSheets/template}
\usepackage{StyleSheets/global}
\usepackage{dsfont}
\usepackage{mdframed}
\usepackage{multicol}

\newif\ifblind\blindtrue    

\begin{document}


\title{Quantum Advantage via Solving Multivariate Polynomials}
\author{Pierre Briaud \thanks{\texttt{pierre@simula.no}} \\ Simula UiB \and Itai Dinur \thanks{\texttt{dinuri@bgu.ac.il}} \\ Ben-Gurion University, Georgetown University  \and Riddhi Ghosal \thanks{\texttt{riddhi@cs.ucla.edu}} \\ UCLA \and Aayush Jain \thanks{\texttt{aayushja@andrew.cmu.edu}} \\CMU \and Paul Lou \thanks{\texttt{pslou@cs.ucla.edu}} \\ UCLA \and Amit Sahai \thanks{\texttt{sahai@cs.ucla.edu}} \\UCLA}
 
\date{}
\maketitle



\begin{abstract}

In this work, we propose a new way to (non-interactively, verifiably) demonstrate quantum advantage by solving the average-case $\mathsf{NP}$ search problem of finding a solution to a system of (underdetermined) constant degree multivariate equations over the finite field $\mathbb{F}_2$ drawn from a specified distribution.
In particular, for any $d \geq 2$, we design a distribution of degree up to $d$ polynomials  $\{p_i(x_1,\ldots,x_n)\}_{i\in [m]}$ for $m<n$ over $\mathbb{F}_2$ for which we show that there is a expected polynomial-time quantum algorithm that provably simultaneously solves $\{p_i(x_1,\ldots,x_n)=y_i\}_{i\in [m]}$ for a random vector $(y_1,\ldots,y_m)$. 
On the other hand, while solutions exist with high probability, we conjecture that for constant $d > 2$, it is classically hard to find one based on a thorough review of existing classical cryptanalysis. Our work thus posits that degree three functions are enough to instantiate the random oracle to obtain non-relativized quantum advantage.

Our approach begins with the breakthrough Yamakawa-Zhandry (FOCS 2022) quantum algorithmic framework. 
In our work, we demonstrate that this quantum algorithmic framework extends to the setting of multivariate polynomial systems.

Our key technical contribution is a new analysis on the Fourier spectra of distributions induced by a general family of distributions over $\mathbb{F}_2$ multivariate polynomials---those that satisfy $2$-wise independence and  shift-invariance.
This family of distributions includes the distribution of uniform random degree at most $d$ polynomials for any constant $d \geq 2$. Our analysis opens up potentially new directions for quantum cryptanalysis of other multivariate systems.

\end{abstract}

\setcounter{page}{0}
\newpage

\section{Introduction}

Since Shor's seminal result~\cite{FOCS:Shor94} over three decades ago showing polynomial-time quantum algorithms for factoring and computing discrete logarithms, researchers have been on the hunt for other average-case search problems in $\NP$ that are efficiently solvable by quantum computers but are believed to be hard classically (``quantum supremacy''~\cite{PreskillQSupremacy}).
Outside the context of $\NP$-search problems, quantum advantage has been identified in several other settings~\cite{clauser1969proposed,FOCS:BCMVV18,brakerski2020simpler,kahanamoku2022classically,kalai2023quantum,jordan2024optimization}.
However, for $\NP$-search problems, examples have remained fairly elusive: solving Pell's equation~\cite{Hal07} and matrix group membership~\cite{STOC:BabBeaSer09}
have efficient quantum algorithms, with both relying on period finding as a crucial tool.
Recently, a breakthrough result of Yamakawa and Zhandry~\cite{FOCS:YamZha22} obtained the first polynomial-time quantum algorithm for an unstructured  \textit{relativized} $\NP$ search problem in the random oracle model that is hard classically, using a fundamentally different approach. Subsequently, the work of Jordan et al.~\cite{jordan2024optimization} used this approach to obtain a quantum algorithm for a combinatorial optimization problem that has an exponential speed-up over known exponential time classical algorithms for that problem, demonstrating a heuristic advantage.
In our work, we will crucially build upon this algorithmic framework, albeit in a very different setting.

We study (conjectured) classically hard average-case NP search problems in the area of solving systems of multivariate polynomial equations. 
Such problems have been studied for decades in the field of multivariate cryptography, where the equations are often quadratic~\cite{C:OngSchSha84,okamoto1986evaluation,EC:MatIma88,C:Patarin95,EC:Patarin96,patarin1997oil,EC:KipPatGou99,C:KipSha99,AC:GouCou00,PKC:DinSch05,ding2005rainbow,chen2008odd,bettale2013cryptanalysis,vates2017key,cabarcas2017key,C:Beullens22} and sometimes cubic~\cite{PQCRYPTO:DinPetWan14,PQCRYPTO:BCEKV18,pepper1,pepper2}. 
Multivariate cryptography has been widely accepted by the cryptographic community as a post-quantum form of cryptography, positing that even quantum algorithms cannot efficiently solve the underlying equations. This belief, however, can be more accurately attributed to the general lack of suitable quantum algorithmic techniques. Indeed, our work provides the first evidence challenging this general belief that classically hard multivariate systems are also quantumly hard, by demonstrating quantum advantage for a family of random multivariate equations with degree at least 3.

In this work, we give an efficient polynomial-time quantum algorithm that provably solves certain families of multivariate polynomial systems, which we will shortly describe. The solutions to these equations generally do not have any nice periodic structure to exploit, nor do we discover any such periodic structure to aid our quantum algorithm. As such, our problem stands apart from previous non-oracle-based quantum advantage results for average-case $\NP$ search problems. We conjecture that this problem is hard to solve using classical methods by analyzing standard solvers for multivariate equations and identifying several features of the equations that could be leveraged for cryptanalysis.

Note that, due to the differences between our equations and those typically considered in multivariate cryptography, we do not expect a direct impact of our work in this particular area. However, our work still offers a fresh perspective on the quantum solving of structured multivariate equations, which may still be of cryptanalytic interest. Indeed, multivariate systems arise in various cryptanalytic contexts, such as solving the Learning With Errors (LWE) problem via the Arora-Ge approach~\cite{arora_ge}, as well as in cryptanalysis attempts on the AES algorithm~\cite{sym1}, among many others. It remains possible that our techniques could prove useful in one of these settings.

\paragraph{The Multivariate System.}
We now describe the simplest concrete multivariate system, among several possible candidates, for which we have an efficient quantum algorithm and for which we conjecture classical hardness.
Since the area of multivariate cryptography has typically focused on the finite field of two elements $\mathsf{GF}(2) = \mathbb{F}_2$ or low-degree field extensions, this setting will be the focus of this paper.

Our multivariate polynomial system over $\bF_2$ consists of two sets of constraints over $n^3$ variables which we denote as $\overline{\mat{x}} = (x_{1, 1}, \ldots, x_{n^2, n})$.
We further group these variables into $\overline{\mat{x}} = (\mbf{x}_1, \ldots, \mbf{x}_{n^2}) \in \bF_2^{n^3}$, where each $\mbf{x}_i \in \bF_2^n$, where $\mbf{x}_i = (x_{i,1}, x_{i,2} \dots, x_{i,n})$.

\begin{enumerate}
    \item The first set of constraints that we impose are $n^2$-many degree three polynomials $\{p_i \}_{i \in [n^2]}$. Each polynomial $p_i$ is uniformly randomly sampled from the set of inhomogeneous degree three polynomials over the variables in $\mbf{x}_i$, that is, the $n$ variables $\{x_{i, j}\}_{j \in [n]}$.
    \item We then impose a set of linear constraints that are described by the dual of a Reed-Solomon\footnote{
    We note that any linear code over $\bF_{2^n}$ that can produce codewords of length $n^2$ and allowing for list decoding from distance $\frac12 + \epsilon$ for any (small) value of $\epsilon$ would suffice for us. We use Reed-Solomon codes just for concreteness.
    } code over an extension field $\mathbb{F}_{2^n}$ with a small constant rate $(1-\alpha)$, where $
    \alpha \in (\frac78,1)$. That is, consider the generating matrix $\mat{H} \in \bF_{2^n}^{(1-\alpha) n^2  \times n^2}$ of this Reed-Solomon code over the extension field $\bF_{2^n}$.
    This generating matrix over $\bF_{2^n}$ can be converted to a generating matrix $\overline{\mat{H}} \in \bF_{2}^{(1-\alpha) n^3 \times n^3}$ over the prime field $\bF_2$ because every $\bF_{2^n}$-linear map can be equivalently expressed as a $\bF_{2}$-linear map.

    Then we impose the linear constraint that $\overline{\mat{H}} \cdot\overline{\mat{x}} = \mbf{0}$, stipulating that any solution should be a codeword of the dual of the Reed-Solomon code.
    Observe that each row of $\overline{\mat{H}}$ defines a linear constraint, imposing (dense) linear dependencies across all the $n^3$ variables. 
    
\end{enumerate}
The computational problem we consider is the task of solving this polynomial system. That is, given the matrix $\overline{\mat{H}}$ and the polynomials $\{p_i\}_{i \in [n^2]}$, compute a vector $\mbf{y} = (\mbf{y}_1, \ldots, \mbf{y}_{n^2}) \in \bF_2^{n^3}$, where each $\mbf{y}_i \in \bF_2^n$,
such that $\overline{\mat{H}} \cdot \mbf{y} = \mbf{0}$ and for all $i \in [n^2]$, $p_i(\mbf{y}_i) = 0$.

\begin{theorem}[Informal Main Theorem]
    There exists an efficient quantum algorithm that given the matrix $\overline{\mbf{H}}$ and the degree-$3$ polynomials $\{p_i\}_{i \in [n^2]}$, computes $\mbf{y} = (\mbf{y}_1, \ldots, \mbf{y}_{n^2}) \in \bF_2^{n^3}$ such that $\overline{\mat{H}} \cdot \mbf{y} = \mbf{0}$ and for all $i \in [n^2]$, $p_i(\mbf{y}_i) = 0$.
\end{theorem}

In fact, our result holds for a larger family of distributions over polynomials $p_i$ above than just random degree-3 or even random degree-$d$ polynomials (with constant $d$). We show that the correctness of our algorithm holds when the polynomials $p_i$ are all independently sampled from \emph{any} fixed distribution that satisfies (1) $2$-wise independence, and (2) shift invariance. See Section~\ref{sec:qeasy} for further discussion and some examples of such distributions.

\paragraph{On Classical Hardness.} We argue that quantum advantage for this problem is achievable when $d \geq 3$, based on our analysis of classical solvers. 

First, we explain why the problem is easy when the first set of constraints contains polynomials of degree $d=2$ instead of $d \geq 3$ by giving a specific polynomial-time algorithm that operates through clever specialization, emphasizing that we are only interested in one of the exponentially many solutions to the multivariate system. We then demonstrate that this algorithm, as well as more general variants reminiscent of algorithms used for the solving of underdetermined multivariate quadratic (MQ) systems~\cite{thomae2021,furue_pqc,hashimoto2023}, become ineffective when $d \geq 3$. 

We complete our analysis by examining standard classical solvers. Among the main algorithms used in cryptanalysis for solving Boolean systems -- specifically, optimized variants of exhaustive search \cite{FES}, 
the polynomial method \cite{lok, bjo, din1, din2}, and Gröbner bases \cite{f4,f5,bettale2009,boolean,bdt} -- we focus on the latter. Indeed, Gröbner basis methods are arguably the most challenging to analyze when applied to structured equations, and thus potentially the most threatening. Since our system is highly underdetermined, it is standard to reduce the number of solutions from exponential to approximately constant before applying these methods. If the initial system were unstructured, or if the resulting overdetermined system had lost some of its initial structure through specialization, we would not be concerned about the cost of Gröbner basis algorithms, as they would then also apply to an unstructured degree-$d$ system with $\mathcal{O}(n^2)$ equations and $\mathcal{O}(n^2)$ variables. However, one might worry that the structure of our equations may contradict this observation or even be leveraged. We provide experimental results and analysis indicating that, to the best of our knowledge, our structure does not enable classical polynomial-time algorithms. By that, we mean that (1) these standard solvers are not expected to run in polynomial time due to the structure of Reed-Solomon code, and (2) it does not seem easy to leverage the efficient decoding procedure to construct other types of efficient solvers.

\paragraph{On Cryptanalytic Impact.} In terms of existing quantum algorithms for solving multivariate polynomial systems that differ from Grover-based exhaustive search \cite{schwabe,pring}, the work of Faugère et al.~\cite{faugere2017fast} gives a faster exponential-time quantum algorithm for solving Boolean quadratic equations, an algorithm that requires on average the evaluation of $\mathcal{O}(2^{0.462n})$ quantum gates. Towards the same goal, Chen and Gao~\cite{chen_gao} proposed a solver that leverages the Harrow-Hassidim-Lloyd (HHL) algorithm \cite{hhl} as a subroutine, with a cost that is polynomial in $n$ and in the condition number of the linear system to which HHL is applied. However, a more recent work \cite{limits} has shown that this condition number is typically extremely high and that the algorithm of~\cite{chen_gao} does not outperform a Grover-based exhaustive search. Note that all these algorithms apply to general multivariate systems. In particular, to the best of our knowledge, no quantum polynomial-time algorithms have been found for possibly more structured ones that are hard to solve classically. Our system provides the first counterexample.

The structure of our equations does not resemble the systems typically encountered in the cryptanalysis of multivariate schemes. They are neither all quadratic, nor similar to the higher-degree equations used in MinRank-based attacks, nor to other types of systems commonly considered in this area. In addition, in the current literature, the equation systems are either overdetermined or mildly underdetermined (e.g. by a constant factor), whereas ours is even more underdetermined (by a polynomial factor). For these reasons, we do not claim any advancement in the quantum cryptanalysis of multivariate cryptography. Nevertheless, our work introduces a new non-generic method of quantumly attacking conjecturably classically hard multivariate systems of equations, offering a fresh perspective and motivation for further work on quantum solving algorithms for structured algebraic systems.

\paragraph{Another Perspective: The Complexity of Instantiating the Yamakawa-Zhandry Random Oracle.}
Another perspective on our work is to think about the problem of un-relativizing Yamakawa-Zhandry: that is, to propose a concrete distribution of simple functions that can be used as a substitute for the random oracle to establish quantum advantage. In particular, our proposal gives a way to replace the Yamakawa-Zhandry constraints that call a random oracle with cubic constraints over $\bF_2$. This is tight, in that substituting the random oracle-based constraint with linear constraints would yield systems of equations that are trivial to solve classically, and we also show non-trivial attacks when the random oracle is replaced with certain family of quadratic constraints. 

\subsection{Technical Overview}\label{sec:tech}
The major technical goal of our paper is to give a quantum algorithm that solves a system of multivariate equations sampled from a distribution satisfying certain properties that we conjecture to be classically hard. For simpler exposition, we will describe our algorithm for a concrete multivariate system. Our result can be generalized to a large family of polynomials. We will briefly mention it at the end of this section.

As described above, we have a polynomial system $\cP$ with $n^2$ many multivariate degree $d>2$ (constant) equations
\[\lbrace p_1,\ldots,p_{n^2}\rbrace \in \bF_2[x_{1,1},\ldots,x_{n^2,n}],\]
and with $(1-\alpha)n^3$ many linear constraints over the $n^3$ variables given by $\overline{\mat{H}}\cdot \overline{\mat{x}}=\mat{0}$, where $\overline{\mat{x}} \coloneqq (x_{1,1},\ldots,x_{n^2,n})$. We often replace $\overline{\mat{x}}$ with simply $\mat{x}$; throughout this paper we will overload notation and interchangeably use $\mat{x}$ to denote an element of $\bF_{2^n}$ and $\bF_2^n$ when it is clear from context.

The linear constraints ensure that if some $\mat{y} \in \bF_2^{n^3}$ satisfies the entire system, then $\mat{y}$ interpreted as $\mat{y}' \in \bF_{2^n}^{n^2}$ must also satisfy a set of linear constraints $\mat{H}\cdot\mat{y}'=\mat{0}$. 
Our choice of $\overline{\mat{H}}$ is such that the corresponding $\mat{H} \in \bF_{2^n}^{((1-\alpha)n^2) \times n^2}$ is a parity-check matrix for a dual Reed-Solomon code $\primal$ over $\bF_{2^n}$ with length $n^2$ and rank $\alpha n^2$.

The breakthrough work of Yamakawa and Zhandry~\cite{FOCS:YamZha22} deals with a related problem in the random oracle model, and crucially uses the properties of random oracles to argue that their quantum algorithm will succeed. Despite these differences from our setting, we (successfully) attempt to carry out their algorithmic framework, albeit with a very different analysis utilizing structural properties of solutions to multivariate polynomials in place of properties of random oracles exploited by Yamakawa and Zhandry. 

We start with Regev's overall quantum algorithmic paradigm~\cite{regev2009lattices} (as did~\cite{FOCS:YamZha22}) for constructing our quantum algorithm.
Our use of a dual Reed-Solomon code constraint is directly lifted from Yamakawa and Zhandry, which fits perfectly with our setting because a dual Reed-Solomon code can be implemented through a system of $\mathbb{F}_2$-linear constraints, as already noted earlier. This fact was not important to Yamakawa-Zhandry but is crucial for us. 

The actual analysis of the algorithm proceeds by, roughly speaking, showing that unique decoding of the Reed-Solomon code will work in quantum superposition. For~\cite{FOCS:YamZha22}, this analysis involved careful and clever analysis of the random oracle.
Our analysis is completely different and based on analyzing properties of degree $d>2$ multivariate polynomials over $\mathbb{F}_2$ (or other prime fields), as we will outline below. 

We begin with a brief overview of how our quantum algorithm works, following the framework of \cite{regev2009lattices,FOCS:YamZha22}:

\paragraph{Quantum Algorithm Overview.} 

Given the explicit description of the generating matrix for $\primal$ and the system of multivariate degree $d>2$ equations $\cP = \lbrace p_1, \ldots, p_{n^2} \rbrace$ in $n^3$ many variables, the algorithm first prepares two uniform superpositions of quantum states:   
\begin{align*}
    \ket{\Psi} &= \sum_{\mat{x} \in \bF_{2^n}^{n^2}} V(\mat{x}) \ket{\mat{x}} \propto \sum_{\mat{x} \in \primal} \ket{\mat{x}}, \\
    \ket{\Phi}& = \sum_{\mat{e} \in \bF_{2^n}^{n^2}} W^\cP(\mat{e}) \ket{\mat{e}} \propto \sum_{\mat{e} = \mat{e}_1 \ldots \mat{e}_{n^2} ~:~ \forall i \in [n^2],~p_i(\mat{e}_i) = 0} \ket{\mat{e}}, 
\end{align*}
where the coefficient functions are defined as
\begin{align*}
    & V(\mat{x})=\begin{cases}
        \frac{1}{\sqrt{ \vert \primal \vert}} \qquad \mat{x} \in \primal\\
        0 \qquad \text{otherwise}
    \end{cases}\\
    & W^{\cP}(\mbf{e}) = \begin{cases}
            2^{-R/2} & \text{if } \forall i \in [n^2], p_i(\mbf{e}_i) = 0\\
            0 & \text{o.w.}
        \end{cases}
\end{align*}
such that $R = \prod_{i \in [n^2]} R_i$ where $R_i$ is the number of roots of polynomial $p_i$.
At a high level, the quantum algorithm aims to take the quantum states $\ket{\Psi}$ and $\ket{\Phi}$ and prepare the pointwise product
\[\sum_{\mat{z} \in \bF_{2^n}^{n^2}} (V \cdot W^{\cP})(\mbf{\mat{z}}) \ket{\mat{z}}.\]
Observe that measuring this pointwise product state is guaranteed to yield a state $\ket{\mat{z}}$ that simultaneously is a codeword of $C$, i.e., $\mat{H} \cdot \mat{z} = \mat{0}$, and is a solution for the degree-$d$ equations, i.e., for all $i \in [n^2]$, we have $p_i(\mat{z}_i) = 0$.  

To produce this pointwise product state, the algorithm first computes the Quantum Fourier Transformation (QFT) of each of $\ket{\Phi}$ and $\ket{\Psi}$ over the extension field $\bF_{2^n}$. 
The QFT of a uniform superposition of codewords is identical to the uniform superposition over the \emph{dual codewords}. This will yield the state 
\[\ket{\hat{\Psi}} \otimes \ket{\hat{\Phi}} =  \sum_{\mat{x} \in \bF_{2^n}^{n^2},\mat{e} \in \bF_{2^n}^{n^2}} \hat{V}(\mat{x})\hat{W}^{\cP}(\mat{e}) \ket{\mat{x}}\ket{\mat{e}},\]
where $\hat{W}^{\cP}(\mbf{e}) = \frac{1}{2^{n^3/2}}\sum_{\mbf{y} \in \bF_{2^n}^{n^2}} W^{\cP}(\mbf{y}) \cdot (-1)^{\trace_{2^n}(\mbf{e} \cdot \mbf{y})}$ and $\hat{V}(\mat{x})=
        \frac{1}{\sqrt{ \vert \dual \vert}} \text{ when } \mat{x} \in \dual, \text{ and }
        0 \text{ otherwise}$.
Then, we compute a unitary addition to obtain $\sum_{\mat{x} \in \dual,\mat{e} \in \bF_{2^n}^{n^2}} \hat{V}(\mat{x}) \hat{W}^{\cP}(\mat{e}) \ket{\mat{x}}\ket{\mat{x}+\mat{e}}$. 
Here, we can view $\mat{e}$ as an error term added to $\mat{x} \in \dual$. 

We can observe at this point that if we were able to surgically remove the first register, the expression $\sum_{\mat{x}, \mat{e} \in \bF_{2^n}^{n^2}} \hat{V}(\mat{x}) \hat{W}^{\cP} (\mat{e}) \ket{\mat{x} + \mat{e}}$ can be equivalently written by the Convolution Theorem as 
\[\sum_{\mat{z} \in \bF_{2^n}^{n^2}}\left( \hat{V} * \hat{W}^{\cP} \right)(\mat{z}) \ket{\mat{z}} = \QFT \sum_{\mat{z} \in \bF_{2^n}^{n^2}} \left(V \cdot W^{\cP}\right) (\mat{z}) \ket{\mat{z}},\]
where the right hand side is exactly our desired state up to a QFT. 
In other words, the critical step is showing that we can uncompute the first register!
To uncompute the first register, the algorithm will apply a decoding algorithm for the Reed-Solomon Code $\dual$ and subtract the output of the decoding algorithm from the first register, i.e.
\begin{align}
\sum_{\mat{x} \in \bF_{2^n}^{n^2},\mat{e} \in \bF_{2^n}^{n^2}} \hat{V}(\mat{x}) \hat{W}^{\cP} (\mat{e}) \ket{\mat{x}-\decode(\mat{x}+\mat{e})}\ket{\mat{x}+\mat{e}} &\approx
\sum_{\mat{x} \in \bF_{2^n}^{n^2},\mat{e} \in \bF_{2^n}^{n^2}} \hat{V}(\mat{x})  \hat{W}^{\cP}(\mat{e}) \ket{\mat{0}}\ket{\mat{x}+\mat{e}} \label{eqn:TechEqn}\\
& = \sum_{\mat{z} \in \bF_{2^n}^{n^2}} (\hat{V} * \hat{W}^{\cP}) (\mat{z}) \ket{\mat{0}} \ket{\mat{z}} \notag.
\end{align}
where the last line follows by the Convolution Theorem (Lemma~\ref{lem:convolution}).
Applying an inverse QFT to the second register yields our desired state $\sum_{\mat{z} \in \bF_{2^n}^{n^2}}\left( V \cdot W^{\cP}\right) (\mat{z}) \ket{\mat{z}}.$

\paragraph{The Technical Heart of Our Work.}
We see now that the key task for proving the algorithm's correctness is proving that there exists a decoding algorithm that succeeds with high probability at uniquely recovering $\mat{x}$ from $\mat{x} + \mat{e}$ where the probability is taken over the distribution of $\mat{e}$ defined by a probability mass function related to the Fourier coefficient function $\hat{W}^{\cP}$.
This begs two questions.
\begin{enumerate}
    \item For what error distributions can we expect to have average-case unique decoding of Reed-Solomon codes?
    \item What is the the error distribution induced by the solutions of degree $d$ multivaiate equations?
\end{enumerate}
The work of Yamakawa-Zhandry~\cite{FOCS:YamZha22} address the first question for an error distribution that is naturally induced by their use of a random oracle.
They proved that if the error distribution is a product distribution over $\bF_{2^n}^m$ of the form $\bar{\cD} = \prod_{i=1}^{m} \cD'$ where $\cD'$ is the distribution  $\bF_{2^n}$ that has $\frac{1}{2}$ probability mass on $\mat{0}$ and is otherwise uniformly distributed over $\bF_{2^n} \setminus \{\mat{0}\}$, then there exists a simple decoding algorithm $\decode$ for any small constant $\varepsilon > 0$ that succeeds with high probability over $\bar{\cD}$:
\begin{enumerate}
    \item List-decode the noisy codeword $\mat{x} + \mat{e}$ to obtain a list $L$ of candidate codewords in $C^\perp$.
    \item If there is a unique codeword $\mat{x}'$ in $L$ such that $\mat{z} - \mat{x}'$ has $\bF_{2^n}$-Hamming weight bounded above by $\left(\frac{1}{2} + \varepsilon\right) n^2$, then output $\mat{x}'$. Otherwise, if there are multiple such candidates, the algorithm fails and returns $\bot$.
\end{enumerate}
This decoding algorithm is natural for this error distribution.
Namely, the first step must use list-decoding because worst-case unique decoding is impossible as there is a significant chance that the relative $\bF_{2^n}$-Hamming weight of the error is at least $\frac{1}{2}$. 
The second step of the algorithm pattern matches the average case error.
Unfortunately, we should not expect such a perfect looking generic distribution $\bar{\cD}$ to capture an error distribution induced by a structured object such as degree $d$ multivariate system of equations.
Nevertheless, after characterizing the error distribution induced by the degree $d$ multivariate system, we will show that we can generalize the class of average-case uniquely decodable distributions to capture it, and moreover the same decoding algorithm $\decode$ will succeed with overwhelming probability for this larger class of error distributions.

\paragraph{The error distribution induced by multivariate degree $d$ polynomials.}
The error distribution we will consider is the error distribution defined by the probability mass function $\bE_{\cP} \left[ \abs{\hat{W}^{\cP}(\cdot) }^2 \right ] : \bF_{2^n}^{n^2} \to [0, 1]$. 
The reason why we focus on this mass function is that if we can show that for this probability mass function an all but negligible amount of the mass is on errors $\mat{e}$ for which for any $\mat{x} \in C^\perp$, $\mat{x} + \mat{e}$ is uniquely decodable to $\mat{x}$, a set of errors that we term
\[\cG \coloneqq \{\mat{e} \in \bF_{2^n}^{n^2}\colon \forall \mat{x} \in \dual, \decode(\mat{x}+\mat{e})=\mat{x}\},\] 
then a simple Markov argument implies that with all but negligible probability over the choice of $\cP$, an all but negligible amount of the mass of the probability mass function $\abs{\hat{W}^{\cP}(\cdot) }^2$ is on $\cG$.

This error distribution with probability mass function $\bE_{\cP} \left[ \abs{\hat{W}^{\cP}(\cdot) }^2 \right ]$ will be a product distribution similar to $\bar{\cD}$.
In particular, it suffices to understand the function $\hat{W}^{p}$ for a single polynomial $p \in \cP$ since these polynomials in the system $\cP$ are each over disjoint blocks of variables and have independently sampled coefficients.

In particular the mass function is as follows: For all $\mat{e}' \in \bF_{2^n}$, 
\[\hat{W}^{p}(\mat{e}') = \frac{1}{2} \left( 1 + (-1)^{p(\mat{z}')}\right) \cdot \frac{1}{\sqrt{R_p}},\] where $R_p$ is the number of roots of $p$. The details of how this expression is derived can be found in the proof of Lemma~\ref{lemma:4wiseErrorDistr}. By a simple calculation (see Lemma~\ref{lemma:4wiseErrorDistr}), we can immediately see that for a single polynomial $p \in \cP$
\[\bE_{p} \left[ \abs{\hat{W}^{\cP}(\mat{0}) }^2 \right ] = \bE_{p} \left[ \frac{R_p}{2^{n}} \right ]  = \frac{1}{2},\]
so that this distribution, like the distribution $\cD'$ previously defined, has $1/2$ probability mass on $\mat{0} \in \bF_{2^n}$. 

For the remaining mass on $\bF_{2^n} \setminus \{ \mat{0} \}$, we need to do some more work. In fact, we will not be able to show that the mass function is uniformly distributed over all $\mat{e} \in \bF_{2^n} \setminus \{\mat{0}\}$. So, we first need to analyze what is a \emph{sufficient} property that this mass function needs to satisfy in order for us to successfully decode with an overwhelming probability. Intuitively, for $\decode$ to succeed, we need to ensure that the probability mass cannot be concentrated on some large hamming weight strings that would break the uniqueness condition required by Step 2 of $\decode$. Therefore it would be sufficient to ensure that there does not exist any $\mat{e} \in \bF_{2^n} \setminus \{\mat{0}\}$ such that $\bE_{p} \left[ \abs{\hat{W}^{\cP}(\mat{0}) }^2 \right]$ is large. In particular, we show that 
\[\forall \mat{e} \in \bF_{2^n}\setminus\{\mat{0}\}, \E_{p} \left[ \abs{\hat{W}^{p}(\mat{e})}^2 \right]  \leq 2^{-\Omega(n)}.\]

To prove this result, the crucial result that we need is that a uniform random degree $d\geq2$ multivariate polynomial is $2$-wise independent.

\begin{lemma}[$2$-wise Independence of Non-linear Polynomials]
    Let $d \geq 2$ be any integer. For any $n \in \bN$, for any two distinct vectors $\mat{x} \neq \mat{y} \in \bF_2^{n}$, for any $a_1, a_2 \in \bF_2$ over the choice of a uniform random inhomogeneous degree $d$ polynomial $p \in \bF_2[X_1,\ldots,X_n]$,
    \[\Pr_{p} \left[ p(\mat{x}) = a_1 \land p(\mat{y}) = a_2 \right ] = \frac{1}{4}.\]
\end{lemma}

\paragraph{Putting Things Together.}

From this analysis, we can conclude that the error distribution is defined by the probability mass function $\bE_{\cP} \left[ \abs{\hat{W}^{\cP}(\cdot) }^2 \right ]$ that is a product distribution of $n^2$-many distributions $\cD_i$ that each have $\frac{1}{2}$ probability mass on $\mat{0}$ and has a mass of no more than $2^{-\Omega(n)}$ at any other point. We show that we will be able to extend the decoder of Yamakawa-Zhandry to work for such distributions.  
This concludes the high-level intuition for how we adapt this algorithmic approach to efficiently solving our family of degree $d$ bounded multivariate systems using a quantum algorithm. 

\paragraph{Some Technical Details.}
To formally prove correctness, however, our analysis will run into technical issues that were previously avoided by the use of the random oracle. 
These are non-trivial obstacles for which we will need to use certain symmetries over the polynomial ring $\bF_{2}[x_1, \ldots, x_n]$. In particular, we also need to rely on the observation that a random multivariate polynomial is \emph{shift-invariant} (see Definition~\ref{def:Invariant}). Specifically we show that for any $\mat{z} \in \bF_{2^n}$, the map $\Pi_z$ that maps a polynomial $p \in \bF_2[X_1,\ldots,X_n]$ to $p' \in \bF_2[X_1,\ldots,X_n]$ such that for all $\mat{x} \in \bF_2^n$, $p'(\mat{x})=p(\mat{x}+\mat{z})$ is a permutation. We refer readers to Section~\ref{section:secondaryTechLemma}, which provides detailed intuition on top of the formal proof. 
\vspace{2em}

\noindent Our result holds for a larger family of polynomials. We show that the correctness of our algorithm holds when polynomials $\cP$ are sampled from \emph{any} distribution that satisfies (1) $2$-wise independence, and (2) shift invariance.

\subsection{Future Work}

\paragraph{Further algorithms using the Regev/Yamakawa-Zhandry framework.}
We hope that our work will help spur the development of other non-relativized efficient quantum algorithms. Natural next steps include varying the linear code used for decoding in the algorithm above, attacking quadratic systems over other prime fields, and attacking higher degree polynomial systems.

\section{Preliminaries}
\label{sec:prelims}
\subsection{Notations}

Lowercase math bold font will be used to denote column vectors $\mbf{x}$ over some finite field $\bF_q = \mathsf{GF}(q)$ where $q = p^e$ for some prime $p$ and positive integer $e$.
Typically, $p = 2$ in our paper.
Uppercase math bold font is used to denote matrices.
Given any two vectors of the same length $\mbf{x}=(x_1,\ldots,x_n), \mbf{y}=(y_1,\ldots,y_n) \in \bF_q^n$, we use $\mbf{x} \cdot \mbf{y}=\sum_{i=1}^n x_iy_i$ to denote the dot product over $\bF_q$. For any matrix $\mat{M}$, $\rk(\mat{M})$ denotes the rank of the matrix.
$[n]$ denotes the set $\{1,2,3,\ldots,n\}$ and $[i..j]$ denotes the set $\{i,i+1,\ldots,j\}$ for $i < j$.

For any vector $\mat{x} \in \bF_q^n$ and any $d \in \bN$, we use the $\hw_{q^d}(\mat{x})$ denotes the $\bF_{q^d}$ Hamming weight of $\mat{x}$. That is, given $\mat{x} \in \bF_q^n$, first split $\mat{x}$ as $\mat{x}_1,\ldots,\mat{x}_{n/d}$ where $\forall i,~\mat{x}_i \in \bF_q^d$. Now the $\hw_{q^d}(\mat{x})$ is the number of non-zero vectors $\mat{x}_i$. The $\bF_{q^d}$ Hamming distance between $\mat{x}_1 \in \bF_q^n$ and $\mat{x}_2 \in \bF_q^n$ is defined to be $\hw_{q^d}(\mat{x}_1-\mat{x}_2)$. 

\begin{remark}[Interpreting $\bF_{2^n}$ elements as $\bF_{2}^n$ elements.]
    Throughout this paper, we abuse notation so that a symbol denoting the extension field element $\mat{z} \in \bF_{2^n}$ will simultaneously refer to a canonical representation of this element as a vector over the prime field $\mat{z} \in \bF_{2}^n$. See Remark~\ref{rem:extinterpret}.
\end{remark}

For any quantum state $\ket{\psi}$, we denote its Euclidean norm as $\norm{\ket{\psi}}$. For any two states $\ket{\psi}, \ket{\phi}$ and any $\varepsilon >0$ , the notation $\ket{\psi} \approx_{\varepsilon} \ket{\phi}$ denotes that $\norm{\ket{\psi} - \ket{\phi}} \leq \varepsilon$.

\subsection{Some Finite Field Results}



\begin{definition}[Traces over Field Extensions~\cite{lidl1997finite}]
\label{def:Tr}
    For $\alpha \in F = \bF_{q^{m}}$ and $K = \bF_q$, the trace $\trace_{F/ K}(\alpha)$ of $\alpha$ over $K$ is given by
    \[\trace_{F/ K}(\alpha) = \alpha + \alpha^{q} + \cdots + \alpha^{q^{m-1}}.\]
    If $K$ is the prime subfield of $F$, then $\trace_{F/ K}(\alpha)$ is called the absolute race of $\alpha$ and simply denoted by $\trace_{F}$.
\end{definition}


\begin{lemma}[Theorem 2.24 in~\cite{lidl1997finite}]
    \label{lemma:traceFieldExt}
  There exists a bijective map $\bm:\bF_{2^n}\rightarrow\bF_{2}^n$ such that
  \begin{itemize}
    \item $\bm(\mbf{0})=0^n$.
      \item For all $\mbf{e} \in \bF_{2^n}$, for all $\mat{z} \in \bF_{2^n}$, we have 
      \[\trace_{2^n}(\mat{e} \cdot \mat{z}) = \langle\bm(\mat{e}_i),\mat{z}\rangle_2,\]
      where $\mat{z}$ is mapped to $\bF_2^{n}$ under the canonical map from $\bF_{2^n} \to \bF_{2}^{n}$. 
  \end{itemize}
  
\end{lemma}

\subsection{Quantum Fourier Transform over Finite Fields}

We import and review the definitions and facts about the quantum Fourier transform over finite fields from the works of~\cite{FOCS:YamZha22,ALGO:dBCW02}.

\begin{definition}[Quantum Fourier Transform]
    For any prime $p$, for any positive integer $m$, for any nonzero linear mapping $\phi : \bF_q \to \bF_p$ where $q = p^m$ so that $\bF_q$ elements are viewed as $m$ dimensional vectors over $\bF_p$, the quantum Fourier transform (QFT) over $\bF_q$ relative to $\phi$ is such that
    \begin{align*}
        \QFT \ket{x} = \frac{1}{q^{1/2}} \sum_{y \in \bF_q} \omega_p^{\phi(xy)} \ket{y}
    \end{align*}
    where $\omega_p = e^{2\pi i / p}.$
\end{definition}

\begin{remark}[QFT Over Arbitrary States]
    This definition is extended to quantum states over multiple qubits by the linearity of $\phi$.
    Namely, for any prime power $q = p^m$ and any vector $\mbf{x} \in \bF_q^n$,
    \begin{align*}
        \QFT \ket{\mbf{x}} = \frac{1}{q^{n/2}} \sum_{\mbf{y} \in \bF_q^n} \omega_p^{\phi(\mbf{x} \cdot \mbf{y})} \ket{\mbf{y}}.
    \end{align*}
    
\end{remark}

\begin{remark}[Trace]
    Throughout this paper, for any prime power $q = p^m$, our linear map $\phi$ will be the absolute trace $\trace_{q}$ over the prime field $\bF_p$.
    In subsequent writing, we will directly replace $\phi$ with $\trace_{q}$.
\end{remark}

\begin{definition}[Post-QFT Fourier Coefficients]
\label{def:fhat}
    For any prime power $q$, for any function $f : \bF_q^n \to \bC$, define
    \begin{align*}
        \hat{f}(\mbf{z}) \coloneqq \frac{1}{q^{n/2}} \sum_{\mbf{x} \in \bF_q^n} f(\mbf{x}) \omega_p^{\trace_q(\mbf{x} \cdot \mbf{z})}.
    \end{align*}
\end{definition}

\begin{definition}[Pointwise Product]
    For any prime power $q$, for any functions $f : \bF_q^n \to \bC$ and $g : \bF_q^n \to \bC$, the pointwise product of $f$ and $g$ is defined to be 
    \begin{align*}
        (f \cdot g)(\mbf{x}) \coloneqq f(\mbf{x}) \cdot g(\mbf{x}).
    \end{align*}
\end{definition}

\begin{definition}[Convolution]
\label{def:star}
    For any prime power $q$, for any functions $f : \bF_q^n \to \bC$ and $g : \bF_q^n \to \bC$, the convolution of $f$ and $g$ is defined to be 
    \begin{align*}
        (f * g)(\mbf{x}) \coloneqq \sum_{\mbf{y} \in \bF_q^n} f(\mbf{y}) \cdot g(\mbf{x} - \mbf{y}).
    \end{align*}
\end{definition}

\begin{lemma}[Parseval's Equality~\cite{FOCS:YamZha22,ALGO:dBCW02}]
\label{lem:parseval}
    For any prime power $q$, for any $f : \bF_q^n \to \bC$,
    \begin{align*}
        \sum_{\mbf{x} \in \bF_q^n} \abs{f(\mbf{x})}^2 =  \sum_{\mbf{x} \in \bF_q^n} \abs{\hat{f}(\mbf{x})}^2 
    \end{align*}
\end{lemma}

\begin{lemma}[Lemma 2.2 in \cite{FOCS:YamZha22}]
\label{lem:yz2.2}
    Let $N = mn$ for $m, n \in \bN$, 
    Suppose that we have $f_i : \bF_q^n \to \bC$ for $i \in [m]$ and $f : \bF_q^N \to \bC$ is defined by 
    \begin{align*}
        f(\mbf{x}) \coloneqq \prod_{i \in [m]} f_i(\mbf{x}_i)
    \end{align*}
    where $\mbf{x} = (\mbf{x}_1, \mbf{x}_2, \ldots, \mbf{x}_m)$. Then, we have
    \begin{align*}
        \hat{f}(\mbf{z}) \coloneqq \prod_{i \in [m]} \hat{f}_i(\mbf{z}_i)
    \end{align*}
    where $\mbf{z} = (\mbf{z}_1, \mbf{z}_2, \ldots, \mbf{z}_m)$.
\end{lemma}

\begin{lemma}[Convolution Theorem, Lemma 2.3 in~\cite{FOCS:YamZha22}]
\label{lem:convolution}
    For any prime power $q$, for functions $f : \bF_q^n \to \bC$, $g : \bF_q^n \to \bC$, and $h : \bF_q^n \to \bC$, the following equations hold.
    \begin{align*}
        \widehat{f \cdot g} &= \frac{1}{q^{n/2}} \left( \hat{f} * \hat{g} \right),\\
        \widehat{f * g} &= q^{n/2} \left( \hat{f} \cdot \hat{g} \right),\\
        \widehat{f \cdot (g * h)} &=  \left( \hat{f} * \left (\hat{g}  \cdot \hat{h} \right)\right).
    \end{align*}
    
\end{lemma}

\subsection{Error Correcting Codes}

\begin{definition}[Linear Codes]
    A linear code $C \subseteq \bF_q^m$ of length $m$ and rank $k$ is a linear subspace of $\bF_q^m$ with dimension $k$.
\end{definition}

\begin{definition}[Dual Codes]
    If $C$ is linear code of length $m$ and rank $k$, then we define the dual code as its orthogonal complement
    \begin{align*}
        C^\perp \coloneqq \left \{ \mbf{z} \in \bF_q^m : \forall \mbf{x} \in C,~\mbf{x} \cdot \mbf{z} = 0 \right \}
    \end{align*}
    $C^\perp$ is a linear code of length $m$ and rank $m-k$ over $\bF_q$. 
\end{definition}

\begin{remark}
    Every linear code $C$ defines a set of constraints on its codewords. To be more specific, there exists a parity-check matrix $\mat{H} \in \bF_q^{(m-k) \times m}$ such that $\forall \mat{z} \in C, \mat{H}\mat{z}=\mat{0}$. Moreover $\mat{H}$ is the generating matrix for the dual code $C^\perp$, i.e., every $\mat{y} \in C^\perp$ lies in the columnspace of $\mat{H}^T$.
\end{remark}

\begin{definition}[List Decoding]
    For any code $C \subseteq \bF_q^m$, list decoding is defined as the following problem: Given some $\mat{x} \in \bF_q^m$ and an error bound $e \in \bN$, the goal is to output a list $L \subseteq C$ that consists exactly of all $\mat{c}$ such that the Hamming distance of $\mat{c}$ and $\mat{x}$ is at most $e$.
\end{definition}

\begin{definition}[Reed-Solomon Codes, see Def. 5.2.1 in~\cite{guruswami2012essential}]
\label{def:rs}
    Let $\bF_q$ be a finite field, and choose $m \in \bN$, $k \in \bN$ such that $k \leq m \leq q$. Fix a sequence of $\mat{\gamma}=(\gamma_1,\ldots,\gamma_m)$ of $m$ distinct elements from $\bF_q$. The encoding function for the Reed-Solomon code with parameters $(m,\mat{\gamma},k)$ over $\bF_q$ denoted by $RS_q[\mat{\gamma},k]:\bF_q^k \rightarrow \bF_q^n$ is defined as follows:

    Map a message $\mat{a}=(a_0,\ldots,a_{k-1})$ with every $a_i \in \bF_q$ to a degree $k-1$ polynomial: $\mat{a} \rightarrow f_\mat{a}(X)$, where \[f_\mat{a}(X)=\sum_{i=0}^{k-1} a_iX^i.\]
    Note that $f_{\mat{a}}(X) \in \bF_q[X]$ is a polynomial of degree at most $k$. The encoding of $\mat{a}$ is the evaluation of $f_{\mat{a}}(X)$ at all $\gamma_i$'s:\[RS_q[\mat{\gamma},k](\mat{a})=(f_{\mat{a}}(\gamma_1),\ldots,f_{\mat{a}}(\gamma_m)).\]
\end{definition}

 
\begin{lemma}[List Decoding of Reed-Solomon Codes~\cite{FOCS:GurSud98}]
\label{lem:rslist}
    There is a classical polynomial time deterministic list decoding algorithm $RSListDecode_{(m, \mat{\gamma}, k)}$ for a Reed-Solomon code $C$ over $\bF_q$ with parameters $(m,\mat{\gamma},k)$ that corrects up to $m-\sqrt{km}$ errors. More precisely, for any $\mat{z} \in \bF_q^m$, $RSListDecode_{(m, \mat{\gamma}, k)}(\mat{z})$ returns the list of all $\mat{x} \in C$ such that $\hw_{q}(\mat{x}-\mat{z}) \leq m-\sqrt{km}$.
\end{lemma}

\begin{definition}[Generalized Reed-Solomon Codes~\cite{guruswami2012essential}]
\label{def:grs}
    A Generalized Reed-Solomon code over $\bF_q$ parameterized by $(m,\mat{\gamma},k,\mat{v})$, where $\mat{v}=(v_1,\ldots,v_m) \in (\bF_q^*)^m$, where $\bF_q^*$ are the nonzero elements from $\bF_q$, is defined similar to a Reed-Solomon code except its encoding function is:
    \[GRS_q[\mat{\gamma},k,\mat{v}](\mat{a})=(v_1 f_{\mat{a}}(\gamma_1),\ldots,v_m  f_{\mat{a}}(\gamma_m)).\]
\end{definition}
\begin{lemma}[Dual Reed-Solomon Codes~\cite{guruswami2012essential}]
\label{lem:dualrs}
    If $C$ is Reed-Solomon code over $\bF_q$ with parameters $(m,\mat{\gamma},k)$, then the dual code $C^\perp$ is a Generalized Reed-Solomon code over $\bF_q$ with parameters $(m,\mat{\gamma},m-k,\mat{v})$ for some $\mat{v} \in (\bF_q^*)^m$.
\end{lemma}

\begin{lemma}[Lemma 4.1~\cite{FOCS:YamZha22}]
\label{lem:yz4.1}
For a linear code $C \subseteq \bF_q^m$, if we define $$f(\mat{x})=\begin{cases}
        \abs{\primal}^{-1/2} & \text{if } \mat{x} \in \primal\\
        0 & \text{o.w.}
    \end{cases},$$ then we have $$\hat{f}(\mbf{x}) = \begin{cases}
        \abs{\dual}^{-1/2} & \text{if } \mbf{x} \in \dual\\
        0 & \text{o.w.}
    \end{cases}.$$
    Here $\hat{f}$ is as per Definition~\ref{def:fhat}.
\end{lemma}

\subsection{Boolean Fourier Analysis}
We now restate and prove Parseval's Theorem in the standard Boolean Fourier analysis context.
We use $\Tilde{f}$ to denote the Fourier coefficient rather than the typical $\widehat{f}$ notation because we have already used $\widehat{f}$ as the Fourier coefficient in the QFT setting, which differs by a normalization factor of $\frac{1}{2^{n/2}}$.
\begin{lemma}[Boolean Parseval's Theorem]
\label{lem:boolparseval}
    Let $p : \{0, 1\}^{n} \to \{0, 1\}$ and define  $P : \{0 , 1\}^{n} \to \{\pm 1\}$ such that $P(\mat{x}) = (-1)^{p(\mat{x})}$. Define $\Tilde{P} :  \{0 , 1\}^{n} \to \bR $ such that $\Tilde{P}(\mat{a}) \coloneqq \E_{\mat{z} \in \bF_2^{n}} \left[  P(\mat{z}) \cdot (-1)^{\langle \mat{a}, \mat{z} \rangle_2 } \right ]$ where $\langle \cdot, \cdot \rangle_2$ denotes $\bF_2$ dot product.
    Then, 
    \[\sum_{\mat{a} \in \bF_2^{n}} \Tilde{P}(\mat{a})^2 = 1.\]
\end{lemma}
\begin{proof}
    For any $\mat{a}, \mat{b} \in \bF_2^{n}$, define the $0/1$ indicator random variable 
    \[\bI_{\mat{a} = \mat{b}} = \begin{cases}
        1 & \text{if } \mat{a} = \mat{b}\\
        0 & \text{o.w.}
    \end{cases} = \E_{\mat{x} \in \bF_2^n} \left [ (-1)^{\langle  \mat{a} \oplus \mat{b}, \mat{x} \rangle_2 } \right ].\]
    Then define $\chi_{\mat{a}}(\mat{b}) \coloneqq (-1)^{\langle \mat{a}, \mat{b} \rangle_2} =  \chi_{\mat{b}}(\mat{a})$.
    \begingroup
    \allowdisplaybreaks
    \begin{align*}
        \sum_{\mat{a} \in \bF_2^n} \Tilde{P}(\mat{a})^2 & = \sum_{\mat{a}, \mat{b} \in \bF_2^{n}} \Tilde{P}(\mat{a}) \Tilde{P}(\mat{b}) \E_{\mat{x} \in \bF_2^n} \left[ \chi_{\mat{a}} (\mat{x}) \chi_{\mat{b}} (\mat{x})\right ]\\
        &=  \E_{\mat{x} \in \bF_2^{n}} \left[ \sum_{\mat{a}, \mat{b}\in \bF_2^{n}}  \Tilde{P} (\mat{a}) \Tilde{P}(\mat{b}) \chi_{\mat{a}} (\mat{x}) \chi_{\mat{b}} (\mat{x})\right ]\\
        &= \E_{ \mat{x} \in \bF_2^{n}} \left[  \left(\sum_{\mat{a} \in \bF_2^n} \Tilde{P} (\mat{a}) \chi_{\mat{a}} (\mat{x}) \right)^2 \right]\\
        &= \E_{ \mat{x} \in \bF_2^{n}} \left[  \left(\sum_{\mat{a} \in \bF_2^n} \E_{\mat{y} \in \bF_2^{n}} \left[  P(\mat{y}) \cdot \chi_{\mat{a}}(\mat{y}) \right ] \chi_{\mat{a}} (\mat{x}) \right)^2 \right]\\
        &=
         \E_{ \mat{x} \in \bF_2^{n}} \left[  \left( \E_{\mat{y} \in \bF_2^{n}} \left[  P(\mat{y}) \cdot \sum_{\mat{a} \in \bF_2^n}\chi_{\mat{x} \oplus \mat{y}}(\mat{a}) \right ] \right)^2 \right]
    \end{align*}
    \endgroup
    If $\mat{x} = \mat{y} \in \bF_2^{n}$, then for all $\mat{a} \in \bF_2^{n}$, $\chi_{\mat{x}  \oplus \mat{y}}(\mat{a}) = 1$ so that $\sum_{\mat{a} \in \bF_2^{n}} \chi_{\mat{x} \oplus \mat{y}}(\mat{a}) = 2^{n}$.
    Otherwise, $\mat{x} \neq \mat{y}$ and the function $\chi_{\mat{x} \oplus \mat{y}}$ is a balanced function, so the sum $\sum_{\mat{a} \in \bF_2^{n}} \chi_{\mat{x} \oplus \mat{y}}(\mat{a}) = 0$.
    Therefore, by expanding the last line above we have that
    \begingroup
    \allowdisplaybreaks
    \begin{align*}
    \sum_{\mat{a} \in \bF_2^n} \Tilde{P}(\mat{a})^2  = \E_{ \mat{x} \in \bF_2^{n}} \left[ P(\mat{x})^2  \right] = 1
    \end{align*}
    \endgroup
    where we observe that $P^2$ is identically $1$.
\end{proof}

\subsection{Useful Results}
\begin{lemma}[Chernoff Bound~\cite{wiki:chernoff}.]
\label{lem:chernoff}
Let $X_1,\ldots,X_n$ be independent Bernoulli random variables. Define $X=\sum_{i=1}^n X_i$ and let $\mu \coloneqq E[X]$. Then for any $\delta>0$, \[\Pr[\vert X - \mu \vert > \delta \mu ] \leq 2e^{-\frac{\delta^2 \mu}{3}}.\]    
\end{lemma}


\section{Our Polynomial System}
\label{sec:mqsystem}

Our polynomial system that is provably quantum easy to invert and that we conjecture to be classical hard is parameterized by any constant $d > 2$,  will consist of $mn$ indeterminates $\{X_{i, j}\}_{i \in [m], j\in [n]}$ and two sets of constraints.

\begin{enumerate}
    \item[] \textbf{Ingredient 1: Uniform Random Constant Degree-$d$ Bounded  Polynomials.} 
    The first set of constraints in our polynomial system are $m$ uniform randomly sampled polynomials of degree at most $d$ over disjoint variables, i.e. 
    \[\cP \coloneqq \{ p_{i} \in \bF_2[X_{i, 1}, \ldots, X_{i, n}] \}_{i \in [m]}.\]

\item[] \textbf{Ingredient 2: Reed-Solomon Code.} Let $\dual$ be a Reed-Solomon code\footnote{We use the notation $\dual$ for the Reed-Solomon code as instead of the more standard $\primal$ for reasons which will be clear in the upcoming sections.} over $\bF_{2^n}$ with parameters $(m, \gamma, (1-\alpha)m)$, for constant $0<\alpha<1$. 

Let $\primal$ be the dual code of $\dual$, i.e., $\primal=(\dual)^\perp$. According to Lemma~\ref{lem:dualrs}, we know that $\primal$ will be a Generalized Reed-Solomon Code over $\bF_{2^n}$ with parameters $(m, \gamma, \alpha m,\mat{v})$ for some $\mat{v} \in \mextension$. 

Observe that the description of $\dual$ gives us the explicit description of $\primal$. Indeed, $\dual$ is given by a system of linear constraints which forms its parity check matrix $\mat{G} \in \bF_{2^n}^{\alpha m \times m}$, thus $\mat{G}$ is also the generating matrix for $\primal$.

Similarly, every $\mat{y} \in \primal$ satisfies a set of linear constraints. In particular, it must be that 
\[\mbf{H}\cdot \mbf{y} = \mbf{0}, \]
where $\mbf{H} \in \bF_{2^n}^{(1- \alpha) m \times m }$ is the parity check matrix for $\primal$ or equivalently the generating matrix for $\dual$.
Note that every $\bF_{2^n}$-linear map $\mbf{H} : \bF_{2^n}^{(1-\alpha) m} \to \bF_{2^n}^{m}$ is equivalent to some $\bF_2$-linear map $\overline{\mbf{H}} : \bF_2^{(1-\alpha)mn} \to \bF_2^{mn}$.

\begin{remark}
    We note that any linear code over $\bF_{2^n}$ that can produce codewords of length $n^2$ and allowing for list decoding from distance $\frac12 + \epsilon$ for any (small) value of $\epsilon$ would suffice for us. We use Reed-Solomon codes just for concreteness.
\end{remark}
\end{enumerate}

\paragraph{Final Polynomial System.} 
Combining the two ingredients defined above, the final polynomial system $\cF$ contains
\begin{enumerate}
    \item $m$-many random $d$-degree bounded polynomials $\cP$.
    \item $(1-\alpha)m$-many linear polynomials defined by $\overline{\mat{H}}$ which is the generating matrix of $\dual$ over $\bF_2$. 
\end{enumerate}
Formally for $\overline{\mbf{x}} = (x_{1, 1}, \ldots, x_{1, n}, x_{2, 1}, \ldots, x_{2, n}, \ldots x_{m, n})$ we have the constraints
\[ 
    \left \{ \begin{array}{c}
         \left \{ p_i(x_{i,1},\ldots,x_{i,n}) = 0 \right\}_{i \in [m]}, \\
         \overline{\mbf{H}} \cdot \overline{\mbf{x}} = \mbf{0}
    \end{array} \right \}.
\]

\begin{remark}
    The above constraints can be equivalently interpreted as $m$-many at most degree $d$ equations on $\alpha m n$ variables by using the linear constraints $\overline{\mbf{H}} \cdot \overline{\mbf{x}} = \mbf{0}$ to backsubstitute for $(1-\alpha) m n$-many variables.
    However, we stress that in our formulation, $\overline{\mbf{H}}$ and $\{p_i\}_{i \in [m]}$ are given to the adversary.
\end{remark}

\paragraph{The Conjectured Classically Hard Problem.} 
Given the system of random polynomials $\cP$ and $\overline{\mbf{H}}$, the task is to find $\mat{y}\coloneqq (\mat{y}_1,\ldots,\mat{y}_m)$ where  for all $i \in [m]$, $\mat{y}_i \in \bF_\q^{n}$, $p_i(\mat{y_i})=0$ \textbf{and} $\overline{\mbf{H}} \cdot \overline{\mbf{x}} = \mbf{0}$.

\begin{remark}[Interpreting $\bF_{2^n}$ elements as $\bF_{2}^n$ elements]
\label{rem:extinterpret}
Note that while the polynomials are evaluated over $\bF_2$, the code $\primal$ is defined over the extension field $\bF_{2^n}$. This will be crucial for the quantum algorithm as we will see later. Observe that every binary string $\mat{y} \in \bF_{2}^n$ can be interpreted in the usual way as a single element over the extension field $\bF_{2^n}$.

That is, $\bF_{2^n}$ is a dimension $n$ vectorspace over $\bF_2$ because $\bF_{2^n} \cong \bF_{2}[x]/q(x)$ for indeterminate $x$ and some irreducible polynomial $q$ of degree $n$ over $\bF_2$.
Therefore, every element $a \in \bF_{2^n}$ can be expressed as a degree $n-1$ polynomial in $z$ given by $a_0 + a_1 z + a_2 z^2 + \cdots + a_{n-1} z^{n-1}$ where $z = x \bmod q(x)$.
Therefore $a \in \bF_{2^n}$ can be represented by a coefficient vector $(a_0, a_1, a_2, \ldots, a_{n-1}) \in \bF_{2}^{n}$. 
\end{remark}

\begin{notation}[Notation: Interpreting $\bF_{2^n}$ elements as $\bF_{2}^n$ elements.]
    Throughout this paper, we abuse notation so that a symbol denoting the extension field element $\mat{z} \in \bF_{2^n}$ will simultaneously refer to a canonical representation of this element as a vector over the prime field $\mat{z} \in \bF_{2}^n$.
\end{notation}


\noindent
\textbf{Parameter Settings:} Let $\lambda$ be the security parameter. Our polynomial system $\cP$ over $\bF_\q$ is parameterized by $(n,m,\alpha,\epsilon)$, where 
\begin{itemize}
    \item $n = \poly(\lambda)$: Number of indeterminates in a single at most degree $d$ polynomial.
    \item $m = \poly(\lambda)$: Number of at most degree $d$ polynomials in $\cP$.
    \item $\alpha$: Positive constant less than $1$. This determines the rank of $\dual$. In particular, $\dual$ has rank $(1 - \alpha) m$.
    \item $\epsilon$: Small positive constant less than $\frac{1}{2}$. This specifies the list-decoding radius of the Reed-Solomon code $\dual$ which enforces the following constraint on $\alpha$ and $\epsilon$:
    \[1 - \sqrt{(1- \alpha)} > \frac{1}{2} + \epsilon.\]
    The quantum algorithm will require for correctness that
    \[ \frac{7}{8} + \frac{3}{4}\varepsilon < \alpha < 1.\]
\end{itemize}
An example of parameter settings with respect to a security parameter $\lambda \in \bN$ for which we have quantum easiness and conjecturable quantum hardness is therefore
\[n = \lambda, m = n^2, \varepsilon = \frac{1}{12}, \alpha = \frac{31}{32}.  \]

\section{Quantum Algorithm}
\label{sec:qeasy}

Our quantum algorithm will in fact succeed even when the polynomials $\cP$ are sampled from any distribution over $n$-variate polynomials over $\bF_2$ that satisfies shift-invariance and standard $2$-wise independence, two properties that we shortly define.
Therefore, we present a correctness theorem for our algorithm for all  distributions $\cD$ that satisfy shift-invariance and standard $2$-wise independence.

We will prove in Section~\ref{sec:constantDegSI4WI} that the distribution of uniform random at most degree $d$ polynomials, for constant $d \in \bN$, satisfy both of the following two properties.

\begin{definition}[Shift-invariant Polynomial Distributions]
    \label{def:Invariant}
    \sloppy
    For $n \in \bN$,  a distribution $\sP$ over $\bF_2[X_1, \ldots, X_n]$  is said to be shift-invariant if for all $\mat{s} \in \bF_2^{n}$, the distribution $\sP_{\mat{s}}$ with a sampling procedure defined by
    \begin{enumerate}
        \item Sample $p \from \sP$.
        \item Output $p'$ where $p'(\mat{x}) = p (\mat{x} + \mat{s})$.
    \end{enumerate}
    is identically distributed as $\sP$.
\end{definition}

\begin{definition}[2-wise Independent Polynomial Distributions]
    \label{def:2wise}
    \sloppy
     For $n \in \bN$ and $k \in \bN$,  a distribution $\sP$ over $\bF_2[X_1, \ldots, X_n]$  is 2-wise independent if
     for any distinct vectors $\mat{x}_1 \neq \mat{x}_2 \in \bF_2^{n}$, for any $a_1 \in \bF_2$ and $a_2 \in \bF_2$,
     \[\Pr_{p \sim \cD} \left[p(\mat{x}_1) = a_1 \land p(\mat{x}_2) = a_2 \right ] = \frac{1}{4}.\]
\end{definition}


\begin{example}[Random High Degree Sparse Polynomials]
   An example of an interesting shift-invariant $2$-wise independent polynomial distribution is the following distribution on $\bF_2[X_1, \ldots, X_n]$ for $n \in \bN$ and for any  $d \in \bN$:
   \begin{enumerate}
       \item Sample a random at most degree $d$ sparse polynomial $p$ so that each coefficient is $1$ with probability $\frac{1}{n^{d- 1}}$ and $0$ otherwise. This polynomial $p$ has $O(n)$ many non-zero coefficients.
       \item Sample a random degree $2$ polynomial $p'$.
       \item Output $p+p'$.
   \end{enumerate} 
   Note here that the sparsity ensures that the representation of this polynomial has a description size bounded by $\poly(n)$. Additionally note that $d$ can have any polynomial relation with $n$ in this setting.
\end{example}

\begin{example}[Random Constant Degree Polynomials]
   Our principal example of an shift-invariant $2$-wise independent polynomial distribution is, as mentioned earlier, when $d = \Theta(1)$, a setting in which we can show that the uniform distribution over at most degree $d$ polynomials in $\bF_2[X_1, \ldots, X_n]$ is shift-invariant and $2$-wise independent.
   See Section~\ref{sec:constantDegSI4WI}.
\end{example}


\begin{theorem}
\label{theorem:main}
    For every constant $\varepsilon \in (0, \frac{1}{6})$, there exists an expected polynomial-time quantum algorithm $\cA$ and a negligible function $\mu : \bN \to [0,1]$, such that for every constant 
    \[ \frac{7}{8} + \frac{3}{4}\varepsilon < \alpha < 1 \]
    for $n = \Theta(\lambda)$, for $m = m(\lambda) = \omega( \log n)$, for $i \in [m]$, for all shift-invariant (Def.~\ref{def:Invariant}) and 2-wise independent (Def.~\ref{def:2wise}) distributions $\sP_i$ with support over $\bF_2[X_{i,1}, \ldots, X_{i,n}]$, for all sufficiently large $\lambda \in \bN$,   the algorithm $\cA$ inverts the polynomial system described by $(\cP, \mat{H})$ with $1 - \mu(\lambda)$ probability over the choice of polynomial system described by $\cP = (p_1, \ldots, p_m)$ where for all $i \in [m]$, $p_i \sim \sP_i$ independently,  and matrix $\mbf{H} \in \bF_{2^{n}}^{(1-\alpha) m \times m}$ is sampled as specified in Section~\ref{sec:mqsystem} with parameter $\alpha$.
\end{theorem}

\begin{proof}
    To prove Theorem~\ref{theorem:main} we introduce the following algorithm, Algorithm~\ref{algo:1}.
    The correctness of Algorithm~\ref{algo:1} is delayed to~\hyperref[proof:correctness]{Proof of Correctness} to allow us to first present the algorithm and introduce several useful technical lemmas for the proof of correctness.
\end{proof}

\begingroup
\begin{algorithm}
    \label{algo:1}
    The algorithm is parameterized by a constant decoding parameter $\varepsilon > 0$.
    The inputs to the algorithm are the description of system of polynomials $\cP = \{p_i \in \bF_2[X_{i, 1}, \ldots, X_{i, n}]\}$ for the indeterminates $(X_{1, 1}, \ldots, X_{1, n}, \ldots, X_{m, 1}, \ldots, X_{m, n})$ sampled as specified in Section~\ref{sec:mqsystem} and the generating matrix for the dual code $\mbf{H} \in \bF_{2^{n}}^{(1-\alpha)m \times m}$.

\label{subsec:alg}
\begin{enumerate}
    \item \label{step:1} For each $i \in [m]$, prepare the uniform superposition of preimages over $\bF_2^{n}$ of $0$ under $p_i$,  
     \[\ket{\phi_i} \propto \sum_{\mbf{e}_i \in \bF_2^{n}~:~ p_i(\mbf{e}_i) = 0}   \ket{\mbf{e}_i}\]
    
    \noindent
    Define the state
 \[\ket{\Phi} \coloneqq \bigotimes_{i =1}^{m} \ket{\phi_i} \propto \sum_{\mbf{e} = (\mbf{e}_1, \ldots, \mbf{e}_m) \in \bF_2^{mn}~:~\forall~i \in [m], p_i(\mbf{e}_i) = 0}\ket{\mbf{e}}.\]
     \item 
    
    Prepare a uniform superposition over all codewords\footnote{Both primal and dual code is over $\bF_{2^n}$. See Remark~\ref{rem:extinterpret}.} for the primal code $\primal$:
    \[\ket{\Psi} \propto \sum_{\mbf{x} \in \primal}  \ket{\mbf{x}},\] 
    In more detail, starting from the uniform superposition $2^{-\alpha m/2} \cdot \sum_{\Tilde{\mbf{x}} \in \bF_{2^n}^{\alpha m}} \ket{\Tilde{\mbf{x}}}$, compute the generating matrix $\mbf{G}$ of the primal code $\primal$ from $\mbf{H}$ and then compute $\ket{\mbf{G} \cdot \Tilde{\mbf{x}}}$ from $\ket{\Tilde{\mbf{x}}}$ to obtain a uniform superposition over all codewords of the primal code, $\ket{\Psi} \propto  \sum_{\mbf{x} \in \primal} \ket{\mbf{x}}$.

    \item Compute $\ket{\hat{\Psi}} = \QFT_{2^n} \ket{\Psi}$ and $\ket{\hat{\Phi}} = \QFT_{2^n} \ket{\Phi}$. The QFTs are taken over the extension field $\bF_{2^n}$.
     

     \item Perform the unitary operation $U_{\mathsf{add}} \left( \ket{\hat{\Psi}} \otimes \ket{\hat{\Phi}} \right)$ which adds the first register to the second register, i.e., 
     
      \[ \ket{\mbf{x}} \ket{\mbf{e}} \xrightarrow{U_{\mathsf{add}}} \ket{\mbf{x}} \ket{\mbf{e} + \mbf{x}}\]
      
      \item Perform the unitary operation $U_{\mathsf{decode}}$ on the quantum state $ U_{\mathsf{add}}  \left( \ket{\hat{\Psi}} \otimes \ket{\hat{\Phi}} \right)$ that first applies the decoding algorithm $\decode$ for $\dual$ to the second register and then subtracts the result from the first register, i.e.,
    \[ \ket{\mbf{x}} \ket{\mbf{e} + \mbf{x}} \xrightarrow{U_{\mathsf{decode}}}\ket{\mbf{x}-\decode(\mat{e}+\mat{x})} \ket{\mbf{e} + \mbf{x}}.\]
    For a constant $0< \epsilon < \frac{1}{2} $, $\decode$ is given by:
\begin{mdframed}
    $\decode(\mat{z})$:
    \begin{enumerate}
        \item On input $\mat{z} \in \mextension$, first run the list decoding algorithm for $\dual$, i.e.\\
        $RSListDecode_{(m, \gamma, (1-\alpha)m)}(\mat{z})$. This gives a polynomial size list of codewords in $\dual$.
        \item If there is a unique $\mat{x}$ in the list such that $\hw_{2^n}(\mat{z}-\mat{x}) \leq \left(\frac{1}{2}+\epsilon\right)m$, then output $\mat{x}$ else output $\mat{0}$.
    \end{enumerate}
\end{mdframed}
    
      \item \label{step:9} Apply the unitary $(I \otimes \QFT_{2^n}^{-1})$ to the resulting state, i.e., compute
      \[(I \otimes \QFT_{2^n}^{-1})\left(U_{\mathsf{decode}} U_{\mathsf{add}}\right)  \left( \ket{\hat{\Psi}} \otimes \ket{\hat{\Phi}} \right).\]
      \item Finally, measure the second register, and output the outcome $\mat{y}^* \in \bF_2^{mn}$.
\end{enumerate}
\end{algorithm}
\endgroup

\paragraph{Notation}
Before proving the correctness and the runtime of the algorithm we introduce the following notation used throughout the proofs.

    Recall the previous notation that $\overline{\mbf{H}} \in \bF_2^{(1-\alpha)mn \times mn}$ is obtained by applying the canonical isomorphism $\bF_{2^n} \cong \bF_2^{n}$ on every entry in the matrix $\mbf{H} \in \bF_{2^{n}}^{(1-\alpha)m \times m}$.
\begin{enumerate}
    \item \textbf{Root Set (Variety) of a polynomial.}    For all $i \in [m]$, let $p_i: \bF_2^n \rightarrow \bF_{2}$ be a polynomial. Define $\sR_i$ as the set of all the roots of $p_i$, i.e., 
\[\sR_i \coloneqq \{\mat{e} \in \bF_2^n \colon p_i(\mat{e}) = 0\}.\]

Next, define the set 
\[\sR \coloneqq \sR_1 \times \sR_2 \times \cdots \times \sR_m.\]

Thus, for every $\mat{e}\coloneqq (\mat{e}_1,\ldots,\mat{e}_m) \in 
\mextension$, we have that $\mat{e} \in \sR$ if $\forall i \in [m], \mat{e}_i \in \sR_i$.

    \item \textbf{Weights used in quantum superposition.} Let $V:\mextension \rightarrow \bC$, $W^{p_i}:\bF_{2^{n}} \rightarrow \bC$, $\forall i \in [m]$, and $W^{\cP}:\mextension \rightarrow \bC$ be functions defined as follows:

    $$V(\mat{x})=\begin{cases}
        \frac{1}{\sqrt{ \vert \primal \vert}} \qquad \mat{x} \in \primal\\
        0 \qquad \text{otherwise}
    \end{cases}$$

     \[W^{p_i}(\mat{e}_i)=\begin{cases}
        \frac{1}{\sqrt{ \vert \sR_i \vert}} & \mat{e}_i \in \sR_i\\
        0 & \text{o.w.}
    \end{cases}
    \implies
    W^{\cP}(\mat{e})=\begin{cases}
        \frac{1}{\sqrt{ \vert \sR \vert}} & \mat{e} \in \sR\\
        0 &  \mathbf{o.w.} \end{cases}
    \]
    Note here that if $p_i$ is identically $1$, i.e. $p_i$ has no roots, then for all $\mat{e}_i \in \bF_2^{n}$, we have $W^{p_i}(\mat{e}_i) = 0$ because $W^{p_i}$ describes the post-partial measurement state of observing $0$ in the first step of the quantum algorithm.

        \item[] Since, $\ket{\Psi}$ is the uniform superposition of all $\mat{x} \in \primal$ and $\ket{\phi_i}$ is the uniform superposition of preimages of $0$ over $\bF_2$ under $p_i$, we have 

 \[\ket{\Psi} = \sum_{\mbf{x} \in \bF_{2^n}^{m}} V(\mbf{x}) \ket{\mbf{x}}\]

    \[\ket{\phi_i} = \sum_{\mbf{e}_i \in \bF_2^{n} }  W^{p_i}(\mbf{e}_i) \ket{\mbf{e}_i}\]

     \[\ket{\Phi} =  \sum_{\mbf{e} = (\mbf{e}_1, \ldots, \mbf{e}_m) \in \bF_{2^n}^{m}} \left( \prod_{i=1}^m W^{p_i}(\mbf{e}_i) \right) \ket{\mbf{e}}=\sum_{\mbf{e} = (\mbf{e}_1, \ldots, \mbf{e}_m) \in \bF_{2^n}^{m}}  W^{\cP}(\mbf{e})  \ket{\mbf{e}}\]
Thus, 
   \[ \ket{\hat{\Psi}} \otimes \ket{\hat{\Phi}} = \sum_{\mbf{x}, \mbf{e} \in \mextension} \hat{V}(\mbf{x}) \cdot \hat{W}^{\cP} (\mbf{e}) \ket{\mbf{x}} \ket{\mbf{e}} \]
    where 
  \[\hat{W}^{\cP}(\mbf{e}) = \frac{1}{2^{nm/2}} \sum_{\mbf{y} \in \mextension} W^{\cP}(\mbf{y}) \cdot (-1)^{\trace_{2^n}(\mbf{e} \cdot \mbf{y})}\]
and where Lemma~\ref{lem:yz4.1} gives
    \[\hat{V}(\mbf{x}) = \begin{cases}
        \abs{\dual}^{-1/2} & \text{if } \mbf{x} \in \dual\\
        0 & \text{o.w.}
    \end{cases}.\] 

\item \textbf{Set of Decodable Error Values.}
\label{def:badSet}
We then introduce the following set definitions. Define $\cG \subseteq \mextension$ as:
\[\cG \coloneqq \{\mat{e} \in \mextension\colon \forall \mat{x} \in C^\perp, \decode(\mat{x}+\mat{e})=x\}.\]
Let $\cB \coloneqq \mextension \setminus \cG$, $\good \coloneqq C^\perp \times \cG$, and 
\[\bad \coloneqq \left(\mextension \times \mextension \right) \setminus \good = \left((\mextension \setminus C^\perp) \times \mextension\right) \cup (C^\perp \times \cB) .\]
Using these set definitions, $\ket{\hat{\Psi}} \otimes \ket{\hat{\Phi}}$ can be written as \[\sum_{\mbf{x}, \mbf{e} \in \mextension} \hat{V}(\mbf{x}) \cdot \hat{W}^{\cP} (\mbf{e}) \ket{\mbf{x}} \ket{\mbf{e}}=\sum_{\mbf{x}, \mbf{e} \in \good} \hat{V}(\mbf{x}) \cdot \hat{W}^{\cP} (\mbf{e}) \ket{\mbf{x}} \ket{\mbf{e}}+\sum_{\mbf{x}, \mbf{e} \in \bad} \hat{V}(\mbf{x}) \cdot \hat{W}^{\cP} (\mbf{e}) \ket{\mbf{x}} \ket{\mbf{e}}.\]

\end{enumerate}


\begin{remark}[Expected Runtime of the Algorithm]
    To observe that the algorithm runs in expected polynomial time, first observe that the QFT can be performed efficiently and the Reed-Solomon list-decoding algorithm for our parameter setting will return a polynomial-sized list.
    The only step that is not clearly efficient is Step~\ref{step:1}, namely the state preparation of the uniform superposition over the roots of a polynomial $p_i$. 

    In more detail, in Step~\ref{step:1}, the algorithm prepares a register that is the uniform superposition over all vectors in $\bF_2^{n}$, it evaluates $p_i$ on the uniform superposition in an auxiliary register, and it measures the auxiliary register.
    If the measurement reads $0$, then the algorithm discards the measured register and the remaining post-measurement state is set to be $\ket{\phi_i}$. If the measurement outcome is not $0$, the algorithm starts this process afresh.
    
    We now formally show that this process terminates in expected constant time. Since $\sP$ is $2$-wise independent, we apply Lemma~\ref{lemma:moments} to see that the number of roots of $p_i \sim \sP_i$, denoted $\abs{\sR_i}$, is tightly concentrated on $2^{n-1}$ with variance $2^{n-2}$.
    Therefore, the variance of $\frac{\abs{\sR_i}}{2^n}$ is $\frac{1}{2^{n+2}}$.
    Then by applying Chebyshev's inequality, we obtain that for all $i \in [m]$, for any constant $\varepsilon > 0$,
    \[\Pr_{p_i \sim \sP_i} \left[ \abs{\frac{\abs{\sR_i}}{2^n} - \frac{1}{2}} \geq \varepsilon\right] \leq  \frac{1}{\varepsilon^2 \cdot 2^{n+2}}.\]
    This implies that the state preparation in Step~\ref{step:1} can be done in expected constant time.
    
\end{remark}

\subsection{Statements of Technical Lemmas}
\label{section:techLemmas}

The proof of Theorem~\ref{theorem:main} will employ two lemmas to show the correctness of Algorithm~\ref{algo:1}.
These two lemmas are similarly proven in~\cite{FOCS:YamZha22} to be true in the context of random oracles.
The second lemma, Lemma~\ref{lemma:secondaryTechnicalLemma}, follows from the first lemma, Lemma~\ref{lemma:mainTechnicalLemma}, in a non-obvious way.
We highlight the proof of the first lemma as our paper's principle technical contribution as it may be of independent interest for future work on quantum algorithms solving multivariate polynomial systems. 

\begin{lemma}
\label{lemma:mainTechnicalLemma}
    Let $\lambda \in \bN$ be the security parameter and let $n = \Theta(\lambda)$ and let $m = \omega(\log n)$.For $i \in [m]$, let $\sP_i$ be any distribution over $\bF_2[X_{i, 1}, \ldots, X_{i, n}]$ that is shift-invariant (Def.~\ref{def:Invariant}) and 2-wise independent (Def.~\ref{def:2wise}). 
    Let $\varepsilon$ be a parameter for the decoding radius, and let $\alpha$ parameterize the rate $1-\alpha$ of a Reed-Solomon code $C^\perp$, and
    let the set $\cB$ be defined as
    \[\cB \coloneqq \{\mat{e} \in \mextension\colon \exists \mat{x} \in C^\perp, \decode(\mat{x}+\mat{e})\neq x\}.\]
    Then, for every constant $\varepsilon \in (0, \frac{1}{6})$, and for every constant $\alpha$ such that
    \[ \frac{7}{8} + \frac{3}{4}\varepsilon < \alpha < 1 \]
    with probability $1 - \negl(n)$ over the choice of $\cP = (p_1, \ldots, p_m)$, in which for all $i \in [m]$, $p_i$ are i.i.d. sampled from $\sP_i$, there exists a negligible function $\mu: \bN \to [0, 1]$ such that for sufficiently large values of $\lambda$, 
    \[\sum_{(\mathbf{x}, \mbf{e}) \in \mathsf{BAD}} \abs{\hat{V}(\mbf{x}) \hat{W}^{\cP}( \mat{e}) }^2  = \sum_{\mat{e} \in \cB} \abs{\hat{W}^{\cP}(\mat{e})}^2 \leq \mu( \lambda).\]
\end{lemma}

\begin{proof}
    We delay the proof to \hyperref[proof:mainTechnicalLemma]{Proof of Main Technical Lemma}.
\end{proof}

\begin{lemma}
\label{lemma:secondaryTechnicalLemma}
    Let $\lambda \in \bN$ be the security parameter and let $n = \Theta(\lambda)$ and let $m = \omega(\log n)$. For $i \in [m]$, let $\sP_i$ be any distribution over $\bF_2[X_{i, 1}, \ldots, X_{i, n}]$  that is shift-invariant (Def.~\ref{def:Invariant}) and 2-wise independent (Def.~\ref{def:2wise}). 
    Let $\varepsilon$ be a parameter for the decoding radius, and let $\alpha$ parameterize the rate $1-\alpha$ of a Reed-Solomon code $C^\perp$, and
    let the set $\cB$ be defined as
    \[\cB \coloneqq \{\mat{e} \in \mextension\colon \exists \mat{x} \in C^\perp, \decode(\mat{x}+\mat{e})\neq x\}.\]
    Then, for every constant $\varepsilon \in (0, \frac{1}{6})$, and for every constant $\alpha$ such that
    \[ \frac{7}{8} + \frac{3}{4}\varepsilon < \alpha < 1 \]
    with probability $1 - \negl(n)$ over the choice of $\cP = (p_1, \ldots, p_m)$ where for all $i \in [m]$, $p_i$ are i.i.d. sampled from $\sP_i$, there exists a negligible function $\delta : \bN \to [0, 1]$ such that for all sufficiently large values of $\lambda$
    \[\sum_{\mat{z} \in \mextension} \left \vert \sum_{(\mat{x},\mat{e}) \in \bad : \mat{x} + \mat{e} = \mat{z}} \hat{V}(\mat{x})\hat{W}^\cP(\mat{e})\right \vert^2 \leq \delta(\lambda).\]
\end{lemma}

\begin{proof}
    We delay the proof to \hyperref[proof:secondaryTechnicalLemma]{Proof of Secondary Technical Lemma}.
\end{proof}

\subsection{Proof of Correctness---Theorem~\ref{theorem:main}}

The overall structure of the proof of correctness of Algorithm~\ref{algo:1} is structurally identical to the proof of correctness in that of Yamakawa-Zhandry~\cite{FOCS:YamZha22}.

\begin{proof}[Proof of Theorem~\ref{theorem:main}]
\label{proof:correctness}

By the definition of the Euclidean norm,
Lemma~\ref{lemma:mainTechnicalLemma} implies that there exists a negligible function $\mu : \bN \to [0, 1]$ such that for all sufficiently large $\lambda \in \bN$,
\[\left\lVert \sum_{(\mbf{x}, \mbf{e}) \in \bad} \hat{V}(\mbf{x}) \cdot \hat{W}^{\cP} (\mbf{e}) \ket{\mbf{x}} \ket{\mbf{e}} \right \rVert \leq \sqrt{\mu(\lambda)}.\]
Since $\bad = \left( \bF_{2^n}^m \times \bF_{2^n}^m\right) \setminus \good$, the above inequality directly implies that
\begin{align}
    \label{eqn:ind1}
    \sum_{(\mat{x}, \mat{e}) \in \bF_{2^n}^m \times \bF_{2^n}^m} \hat{V}(\mat{x}) \hat{W}(\mat{e}) \ket{\mat{x}} \ket{\mat{e}} \approx_{\sqrt{\mu}} \sum_{(\mat{x}, \mat{e}) \in \mathsf{GOOD}}  \hat{V}(\mat{x}) \hat{W}(\mat{e}) \ket{\mat{x}} \ket{\mat{e}}.
\end{align}
Similarly, Lemma~\ref{lemma:secondaryTechnicalLemma} implies that there exists a negligible function $\delta : \bN \to [0, 1]$ such that for all sufficiently large $\lambda \in \bN$,
\[\left \lVert \sum_{(\mbf{x}, \mbf{e}) \in \bad} \hat{V}(\mbf{x}) \cdot \hat{W}^{\cP} (\mbf{e}) \ket{\mbf{x+e}} \right \rVert \leq \sqrt{\delta(\lambda)}\]
which implies
\begin{align}
    \label{eqn:ind2}
    \sum_{(\mat{x}, \mat{e}) \in \bF_{2^n}^m \times \bF_{2^n}^m} \hat{V}(\mat{x}) \hat{W}(\mat{e}) \ket{\mat{x} + \mat{e}} \approx_{\sqrt{\delta}} \sum_{(\mat{x}, \mat{e}) \in \mathsf{GOOD}}  \hat{V}(\mat{x}) \hat{W}(\mat{e}) \ket{\mat{x} + \mat{e}}.
\end{align}
Equation~\ref{eqn:ind1} and Equation~\ref{eqn:ind2} allow us to observe the following sequence of relations:
\begin{align*}
    \left(U_{\mathsf{decode}} U_{\mathsf{add}}\right)  \left( \ket{\hat{\Psi}} \otimes \ket{\hat{\Phi}} \right)
    &=  \left(U_{\mathsf{decode}} U_{\mathsf{add}}\right)  \left( \sum_{(\mbf{x}, \mbf{e}) \in \mextension \times \mextension} \hat{V}(\mbf{x}) \hat{W}^{\cP} (\mbf{e}) \ket{\mbf{x}} \ket{\mbf{e}} \right)\\
    &\approx_{\sqrt{\mu}} \left(U_{\mathsf{decode}} U_{\mathsf{add}}\right)  \left( \sum_{(\mbf{x}, \mbf{e}) \in \good} \hat{V}(\mbf{x}) \hat{W}^{\cP} (\mbf{e}) \ket{\mbf{x}} \ket{\mbf{e}} \right) \tag{By Equation~\ref{eqn:ind1}.}\\
    &= \sum_{(\mbf{x}, \mbf{e}) \in \good} \hat{V}(\mbf{x}) \hat{W}^{\cP} (\mat{e}) \ket{\mat{0}} \ket{\mbf{x} + \mbf{e}} \tag{By definition of $\good$.}\\
    &\approx_{\sqrt{\delta}} \sum_{(\mbf{x}, \mbf{e}) \in \mextension \times \mextension } \hat{V}(\mbf{x}) \hat{W}^{\cP} (\mat{e}) \ket{\mat{0}} \ket{\mbf{x} + \mbf{e}} \tag{By Equation~\ref{eqn:ind2}.}\\
    &= \sum_{\mat{z} \in \mextension} (\hat{V} * \hat{W}^{\cP}) (\mat{z}) \ket{\mat{0}} \ket{\mat{z}} \tag{By definition of convolution.} \\
    &= 2^{mn/2} \sum_{\mat{z} \in \mextension} (\widehat{V \cdot W^{\cP}})(\mat{z}) \ket{\mat{0}} \ket{\mat{z}} \tag{By Lemma~\ref{lem:convolution}.}\\
    &= (I \otimes \QFT_{2^n}) 2^{mn/2} \sum_{\mat{z}\in \bF_{2^n}^m}   (V\cdot W^{\cP})(\mat{z})  \ket{\mbf{0}} \ket{\mbf{z}} \tag{By definition of QFT.}.
\end{align*}
Therefore in Step~\ref{step:9}, the algorithm obtains the following state:
\begin{align*}
    (I \otimes \QFT_{2^n}^{-1})\left(U_{\mathsf{decode}} U_{\mathsf{add}}\right)  \left( \ket{\hat{\Psi}} \otimes \ket{\hat{\Phi}} \right) \approx_{\sqrt{\mu} + \sqrt{\delta}} 2^{mn/2} \sum_{\mat{z}\in \bF_{2^n}^m}   (V\cdot W^{\cP})(\mat{z})  \ket{\mbf{0}} \ket{\mbf{z}}.
\end{align*}
By the definition of the $V$ and $W^{\cP}$, a measurement of the second register on the right hand side is guaranteed to observe a value $\mat{z} = \mat{z}_1 \mat{z}_2 \cdots \mat{z}_{m} \in \bF_{2^{n}}^{m}$ that simultaneously satisfies that $\mat{z} \cdot \mat{H} = \mat{0}$ and for all $i \in [m]$, $p_i(\mat{z}_i) = 0$.
By the indistinguishability of the left and right hand side, with at least $1 - (\mu + \delta)(\lambda)$ probability a measurement of the second register on the left hand side will also satisfy these constraints.
This concludes our proof of correctness.
\end{proof}

\subsection{\texorpdfstring{Proof of Secondary Technical Lemma: Lemma~\ref{lemma:secondaryTechnicalLemma}}{Proof of Secondary Technical Lemma}}
\label{section:secondaryTechLemma}
We now prove the secondary technical lemma, Lemma~\ref{lemma:secondaryTechnicalLemma}, by assuming the main technical lemma,  Lemma~\ref{lemma:mainTechnicalLemma}. 
A reader may choose to skip the following exposition on the intuition behind the proof and instead jump directly to the \hyperref[proof:secondaryTechnicalLemma]{Proof of Secondary Technical Lemma}.

For intuition, recall that  Lemma~\ref{lemma:mainTechnicalLemma} bounds a sum of norm values squared given by 
\[\sum_{(\mathbf{x}, \mbf{e}) \in \mathsf{BAD}} \abs{\hat{V}(\mbf{x}) \hat{W}^{\cP}( \mat{e}) }^2  = \sum_{\mat{e} \in \cB} \abs{\hat{W}^{\cP}(\mat{e})}^2.\]
The obvious idea to relate the two lemmas is to attempt to apply a triangle inequality on the expression in Lemma~\ref{lemma:secondaryTechnicalLemma},
\[\sum_{\mat{z} \in \mextension} \left \vert \sum_{(\mat{x},\mat{e}) \in \bad : \mat{x} + \mat{e} = \mat{z}} \hat{V}(\mat{x})\hat{W}^\cP(\mat{e})\right \vert^2.\] 
Unfortunately, the triangle inequality is false for upper bounding the norm squared of a sum by the sum of norms squared.
Therefore, to semantically match the statements of the two lemmas, we aim to remove the inner summation.

A natural way to remove the summation is to introduce a convolution operator, which by its definition would absorb the inner summation.
Namely, observe that our starting expression can be equivalently rewritten as
\[\sum_{\mat{z} \in \mextension} \left \vert \sum_{(\mat{x},\mat{z} - \mat{x}) \in \bad} \hat{V}(\mat{x})\hat{W}^\cP(\mat{z} - \mat{x})\right \vert^2.\]
The inner expression already resembles a convolution!
To equivalently rewrite the inner expression as a convolution, however, we require the inner sum to range over all $\mat{x}$ values in $\mextension$.
Towards this end, observe that since $\hat{V}$ takes the value $0$ on all vectors in $\mextension \setminus C^\perp$, we can equivalently rewrite the above as
\[\sum_{\mat{z} \in \mextension} \left \vert \sum_{(\mat{x},\mat{z} - \mat{x}) \in \mextension \times \cB} \hat{V}(\mat{x})\hat{W}^\cP(\mat{z} - \mat{x})\right \vert^2.\]
The inner summation now nearly looks like a convolution with the exception of the required condition on $\mat{z} - \mat{x} \in \cB$.
To address this, we apply the exactly same idea as we did with $\hat{V}$, namely we introduce an indicator function $\hat{B} : \mextension \to \bC$ that is exactly $1$ if its input is in the set of values $\cB$ defined above to be the set of ``bad" error values and $0$ otherwise, i.e.
    \[\hat{B} (\mat{e}) = \begin{cases}
        1 & \text{if $\mat{e} \in \cB$.}\\
        0 & \text{o.w.}
    \end{cases}.\]
Finally, we can equivalently rewrite the above expression as
\[\sum_{\mat{z} \in \mextension} \left \vert \sum_{\mat{x} \in \mextension} \hat{V}(\mat{x}) \cdot (\hat{B} \cdot \hat{W}^\cP) (\mat{z} - \mat{x})\right \vert^2 = \sum_{\mat{z} \in \mextension} \left \vert \left(\hat{V} * (\hat{B} \cdot \hat{W}^{\cP})\right) (\mat{z})\right \vert^2 .\]
How do we relate this above expression to that expression in Lemma~\ref{lemma:mainTechnicalLemma}?

Recall that the main expression in Lemma~\ref{lemma:mainTechnicalLemma} can be written equivalently as $\sum_{\mat{e} \in \cB} \abs{\hat{W}^{\cP}(\mat{e})}^2$, an expression that notably has no dependence on $\hat{V}$. 
The above convolution, however, tangles $\hat{V}$ in a non-separable way.
If the inner expression $\hat{V} * (\hat{B} \cdot \hat{W}^{\cP})$ were instead a pointwise product, we could have multiplicatively factored $\hat{V}$ out of the summation since the value of $\hat{V}$ is an indicator function whose value is completely known to us. 
Fortunately, transforming a convolution into a pointwise product is exactly what the Convolution Theorem (see Lemma~\ref{lem:convolution}) does.
Namely, applying the Convolution Theorem twice gives us 
\[\hat{V} * (\hat{B} \cdot \hat{W}^{\cP}) = \widehat{V \cdot (B * W^{\cP})}.\]
We still cannot separate out $V$ since there is a hat sitting on the entire expression. 
Fortunately the summation over $\mat{z}$ is over the entirety of $\mextension$, so we can appeal to Parseval's Theorem (see Lemma~\ref{lem:parseval}) to remove the hat and obtain the following equivalent expression to the original statement of Lemma~\ref{lemma:secondaryTechnicalLemma},
\[\sum_{\mat{z} \in \mextension} \left \vert \left( V \cdot (B * W^{\cP}) (\mat{z}) \right) \right \vert^2,\]
allowing us to finally pull out $V$ from the expression to obtain
\[ \frac{1}{\abs{C}} \cdot \sum_{\mat{z} \in C} \left \vert (B * W^{\cP}) (\mat{z}) \right \vert^2.\]
Unfortunately, the function $B$, which is the dual function to our indicator function $\hat{B}$, is not readily interpretable, nor is its convolution with $W^{\cP}$.
It seems that we have hit the end of the road as we cannot apply Parseval's with the summation on $\mat{z}$ being only over the code $C$.
If the summation were instead over $\mextension$, Parseval's Theorem would be our ticket back to $\hat{B} \cdot \hat{W}^{\cP}$.
Can we find a ticket back? 

What allows us to return to the desired expression is the following symmetry on the set of all $n$-variate $\bF_2$ polynomials.


\begin{lemma}    \label{lemma:convolutionEquality}
        For any $\mbf{z},\mbf{z}' \in \mextension$, for any $m$-many polynomials  $\cP = (p_1, \ldots, p_m)$ sampled such that $\forall i \in [m],~p_i \sim \sP_i$ where $\sP_i$ is shift-invariant (Def.~\ref{def:Invariant}), we have that \[\E_\cP\left[\abs{\left( B * W^\cP\right)(\mbf{z})}^2\right]=\E_\cP\left[\abs{\left( B * W^\cP\right)(\mbf{z'})}^2\right].\]
\end{lemma}
\begin{proof} It suffices to show that the left hand side expression is independent of $\mat{z}$. Here we will use the fact that the distribution $\sP_i$ is shift-invariant and therefore shift-invariant. 
    \begin{align*}
       &\E_{\cP \sim \prod_{i=1}^{m} \sP_i } \left[\abs{\left( B * W^{\cP}\right)(\mbf{z})}^2\right]\\
       =& \E_{\cP \sim \prod_{i=1}^{m} \sP_i}\left[\abs{\sum_{\mat{x} \in \mextension} B(\mat{x}) W^\cP(\mbf{z}-\mat{x})}^2\right]\\
       =&  \E_{\cP \sim \prod_{i=1}^{m} \sP_{i, \mat{z}_i}}\left[\abs{\sum_{\mat{x} \in \mextension} B(\mat{x}) W^{\cP}(\mat{x})}^2 \right]\\
       =& \E_{\cP \sim \prod_{i=1}^{m} \sP_i}\left[\abs{\sum_{\mat{x} \in \mextension} B(\mat{x}) W^{\cP}(\mat{x})}^2\right]. \tag{By the shift invariance property.}
    \end{align*} 
\end{proof}
Lemma~\ref{lemma:convolutionEquality} shows us that taking an expectation over the expression in Lemma~\ref{lemma:secondaryTechnicalLemma} allows the observed symmetry to introduce a summation over the entirety of $\mextension$, enabling us to apply Parseval's Theorem and return from $B * W^{\cP}$ to the known quantity of $\hat{B} \cdot \hat{W}^{\cP}$.
Namely, for any $\mat{z} \in \mextension$,
\begin{align*}
      \E_{\cP} \left [\abs{\left( B * W^\cP\right)(\mbf{z})}^2\right] &=  \frac{1}{2^{nm}} \cdot \E_{\cP} \left [ \sum_{\mat{z'} \in \mextension} \abs{\left( B * W^\cP\right)(\mbf{z'})}^2\right]\\
      &= \E_{\cP} \left [ \sum_{\mat{z'} \in \mextension} \abs{\left( \hat{B} \cdot \hat{W}^\cP\right)(\mbf{z'})}^2\right] \tag{By the Convolution Theorem.}\\
      &=  \E_{\cP} \left [ \sum_{\mat{z'} \in \cB} \abs{ \hat{W}^\cP(\mbf{z'})}^2\right].
\end{align*}
Observe that $\sum_{\mat{z'} \in \cB} \abs{ \hat{W}^\cP(\mbf{z'})}^2$ is exactly what Lemma~\ref{lemma:mainTechnicalLemma} upper bounds, so we have established a relation between the two lemmas.
We now provide the formal proof.
\begin{proof}[Proof of Lemma~\ref{lemma:secondaryTechnicalLemma}]
    \label{proof:secondaryTechnicalLemma}

    By Markov's inequality, to prove Lemma~\ref{lemma:secondaryTechnicalLemma} it suffices to prove the existence of a negligible function $\delta : \bN \to [0, 1]$ such that
\begin{align*}
\E_\cP\left[\sum_{\mat{z} \in \mextension} \left \vert \sum_{(\mat{x},\mat{z} - \mat{x}) \in \bad} \hat{V}(\mat{x})\hat{W}^\cP(\mat{z} - \mat{x})\right \vert^2\right] \leq \delta(\lambda).    
\end{align*}
    We now rewrite the expression inside the expectation in a sequence of steps.
    Recalling the definition of the indicator function $\hat{B} :\bF_{2^n}^m \to \bC$ introduced above, we can rewrite the main expression in Lemma~\ref{lemma:secondaryTechnicalLemma}
\begingroup
    \allowdisplaybreaks
    \begin{align*}
        &\sum_{\mat{z} \in \mextension} \left \vert \sum_{(\mat{x},\mat{e}) \in \bad : \mat{x} + \mat{e} = \mat{z}} \hat{V}(\mat{x})\hat{W}^\cP(\mat{e})\right \vert^2\\
        &\qquad = \sum_{\mat{z} \in \mextension} \left \vert \sum_{(\mat{x},\mat{z} - \mat{x}) \in \bad} \hat{V}(\mat{x})\hat{W}^\cP(\mat{z} - \mat{x})\right \vert^2\\
        &\qquad = \sum_{\mat{z} \in \mextension} \left \vert \sum_{(\mat{x},\mat{z} - \mat{x}) \in \bF_{2^{n}}^{m} \times  \cB} \hat{V}(\mat{x})\hat{W}^\cP(\mat{z} - \mat{x})\right \vert^2 \tag{Since $\hat{V}(\mat{x}) = 0$ for all $\mat{x} \notin C^\perp$.}\\
        &\qquad = \sum_{\mat{z} \in \mextension} \left \vert \sum_{\mat{x} \in \mextension} \hat{V}(\mat{x}) \cdot  \left( \hat{B} \cdot \hat{W}^\cP\right)(\mat{z} - \mat{x})\right \vert^2 \tag{By definition of $\hat{B}$.}\\
        &\qquad = \sum_{\mat{z} \in \mextension} \left \vert  \left( \hat{V} * (\hat{B} \cdot \hat{W}^{\cP})\right) ( \mat{z}) \right \vert^2 \tag{By definition of convolution.}\\
        &\qquad = \sum_{\mat{z} \in \mextension} \left \vert  \left( \widehat{ V \cdot (B * W^{\cP})}\right)  ( \mat{z})\right \vert^2 \tag{Two applications of the Convolution Theorem.}\\
        &\qquad = \sum_{\mat{z} \in \mextension} \left \vert  \left(V \cdot (B * W^{\cP})\right)  ( \mat{z})\right \vert^2 \tag{By Parseval's Theorem.}\\
        &\qquad = \frac{1}{\abs{C}} \cdot \sum_{\mat{z} \in C} \left \vert (B * W^{\cP}) ( \mat{z}) \right \vert^2 \tag{By definition of $V$.}\\
    \end{align*}
\endgroup
Therefore, 
\begingroup
     \allowdisplaybreaks
     \begin{align*}
        &\E_\cP\left[\sum_{\mat{z} \in \mextension} \left \vert \sum_{(\mat{x},\mat{z} - \mat{x}) \in \bad} \hat{V}(\mat{x})\hat{W}^\cP(\mat{z} - \mat{x})\right \vert^2\right]\\
        &\qquad =   \frac{1}{\abs{C}} \cdot  \E_\cP\left[\sum_{\mat{z} \in C} \left \vert  (B * W^{\cP}) ( \mat{z}) \right \vert^2\right] \tag{By substituting the above sequence.}\\
        &\qquad = \frac{1}{\abs{C}}\cdot  \sum_{\mat{z} \in C}  \E_\cP\left[ \left \vert  (B * W^{\cP}) ( \mat{z}) \right \vert^2\right] \tag{By the linearity of expectation.}\\
        &\qquad = \frac{1}{\abs{C} \cdot 2^{nm}} \cdot \sum_{\mat{z} \in C} \E_{\cP} \left [ \sum_{\mat{z'} \in \mextension} \abs{\left( B * W^\cP\right)(\mbf{z'})}^2\right] \tag{By Lemma~\ref{lemma:convolutionEquality}.}\\
        &\qquad = \frac{1}{2^{nm}} \cdot \E_{\cP} \left [ \sum_{\mat{z'} \in \mextension} \abs{\left( B * W^\cP\right)(\mbf{z'})}^2\right] \\
        &\qquad = \E_{\cP} \left [ \sum_{\mat{z'} \in \mextension} \abs{\left( \hat{B} \cdot \hat{W}^\cP\right)(\mbf{z'})}^2\right] \tag{By the Convolution Theorem.}\\
        &\qquad =\E_{\cP} \left [ \sum_{\mat{z'} \in \cB} \abs{ \hat{W}^\cP(\mbf{z'})}^2\right] \tag{By definition of $\hat{B}$.}\\
        &\qquad \leq \mu(\lambda). \tag{By Lemma~\ref{lemma:mainTechnicalLemma}.}
    \end{align*}
\endgroup
\noindent
Letting $\delta(\lambda) = \mu(\lambda)$ concludes the proof. 
\end{proof}

\subsection{\texorpdfstring{Main Technical Lemma Proof: Lemma~\ref{lemma:mainTechnicalLemma}}{Proof of the Main Technical Lemma}}

The proof of Lemma~\ref{lemma:mainTechnicalLemma} is our main technical contribution and the supporting lemmas used and proven in the process may be of independent interest for how to apply quantum Fourier transformations to solve multivariate polynomial systems.

\paragraph{Interpretation of Lemma~\ref{lemma:mainTechnicalLemma}.}
The use of Lemma~\ref{lemma:mainTechnicalLemma} in the proof of correctness is to show that applying $\decode$ in superposition to uncompute the first register will succeed on an overwhelming measure of the terms in the superposition. 
Therefore,
Lemma~\ref{lemma:mainTechnicalLemma} is best framed by viewing the function $\hat{W}^{\cP} : \mextension \to \bC$ as inducing a probability distribution defined by the probability mass function $\E_{\cP} \left[ \abs{\hat{W}^{\cP}(\cdot)}^2  \right ] : \mextension \to [0,1]$ over the error values $\mat{e} = \mat{e}_1 \mat{e}_2 \cdots \mat{e}_m \in \mextension$.
This is a distribution that notably does not depend on any specific choice of $\cP$.
Lemma~\ref{lemma:mainTechnicalLemma} states exactly that 
random errors sampled from this 
distribution are, with all but negligible probability, \emph{uniquely} decodable under the classical decoding algorithm $\decode$.
Recall that our decoding algorithm $\decode$ proceeds in two natural steps:
\begin{enumerate}
    \item List-decode the noisy codeword $\mat{z} \coloneqq \mat{x} + \mat{e}$ to obtain a list $L$ of candidate codewords in $C^\perp$.
    \item If there is a unique codeword $\mat{x}'$ in $L$ such that $\mat{z} - \mat{x}'$ looks like a ``reasonable" error, i.e. the $\bF_{2^n}$-Hamming weight is bounded above by $\left(\frac{1}{2} + \varepsilon\right) m$, then output $\mat{x}'$. Otherwise, if there are multiple such candidates, the algorithm fails and returns $\bot$.
\end{enumerate}
List-decoding, rather than unique decoding, must be performed as the worst case $\bF_{2^n}$-hamming weight is well beyond the unique decoding radius.
For any generator $\gamma$ and $\alpha \in (0, 1)$, one can uniquely decode noisy $(m, \gamma, (1-\alpha)m)$-Reed-Solomon codewords with worst-case errors of up to $\left \lfloor \frac{\alpha m}{2}\right \rfloor$ $\bF_{2^n}$-Hamming weight.
However, we can only guarantee that this error distribution outputs errors that have  at most  $(\frac{1}{2} + \varepsilon) m$ $\bF_{2^n}$-Hamming weight with high probability, for some small constant $\varepsilon > 0$.
Therefore, successfully running unique decoding instead of list decoding as the first step would impose the requirement that $\alpha \geq 1 + 2\varepsilon$, a nonsensical setting of the rate parameter $\alpha$. 
Moreover, since the error's relative $\bF_2$-hamming weight may be well over $\frac{1}{2}$, we must perform this list-decoding process over the field extension, enlarging our alphabet size to avoid the Johnson bound.
Once we've obtained a list of candidates from the list decoding step, we can use a straightforward counting argument to argue that with high probability, there exists a unique candidate in the list that satisfies our Hamming weight constraint.
In other words, this argument gives average-case unique decoding for this type of error distribution.


\subsubsection{Average-case Unique Decoding of Reed-Solomon Codes}

We now generalize the proof of Item 2 in Lemma~4.2 in~\cite{FOCS:YamZha22} to characterize a more general distribution of error distribution for which $\decode$ can successfully recover the original codeword. 
This may be of interest for extending the application of the Yamakawa-Zhandry algorithm~\cite{FOCS:YamZha22} to other instantiations of the random oracle beyond an MQ polynomial system.

Previously, the condition in the analogous theorem in~\cite{FOCS:YamZha22} considers a distribution in which $\mbf{0} \in \bF_2^{n}$ has probability mass $\frac{1}{2}$ and is otherwise uniform over $\bF_2^{n}$.
We observe that we can relax this condition to stipulating that the remaining $\frac{1}{2}$ probability mass is not concentrated on any particular element in $\bF_2^{n} \setminus \{ \mbf{0}\}$.

\begin{lemma}[Generalization of Item 2 in Lemma~4.2 in~\cite{FOCS:YamZha22}]
    \label{lemma:decodableBurstErrors}
    Let $n = \Theta(\lambda)$ and $m =\omega(1)$.
    Let $\cD$ by any error distribution described by a product distribution $\cD = \prod_{i \in [m]} \cD_i$ that satisfies the following property:
    \begin{itemize}

        \item 
        For $i \in [m]$, let $\cD_i$ be a distribution over $\bF_2^n$ that takes $\mbf{0} \in \bF_2^{n}$ with probability $\frac{1}{2}$ and otherwise takes any element in the support with probability at most $2^{-c_i n}$ for some constant $c_i > 0$. Let $c_{\min} \coloneqq \min_{i \in [m]} c_i$.

    \end{itemize}
    For any arbitrarily small constant $\varepsilon > 0$, for any $\alpha > \max \Bigg(  \left(\frac{3}{4} + \frac{3 \varepsilon}{2} \right),~ \left(1 - \frac{c_{\min}}{4} + \varepsilon \cdot \left (1 - \frac{c_{\min}}{2} \right) \right)\Bigg)$, for any generator $\gamma \in \bF_{2^n}^* \cong \bZ_{2^{n-1}}$, let $ \{ C^\perp_{\lambda}\} $ be a  $(m, \gamma, (1-\alpha)m)$-Reed-Solomon code family
    over the finite field $\bF_{2^n}^{m}$.
    Then, there exists an efficient unique decoding algorithm $\decode$ for $C^\perp$ for $\cD$, i.e.
    \begin{align*}
        \Pr_{\mat{e} = \mat{e}_1\cdots \mat{e}_m~\sim~\prod_{i \in [m]} \cD_i} \left[ \forall~\mat{x} \in C^{\perp}, \decode(\mat{x} + \mat{e}) = \mat{x} \right] \geq 1 - 2^{-\Omega(\lambda)}.
    \end{align*}
\end{lemma}
\begin{proof}
First, since each distribution $\cD_i$ is independent and takes on the value $\mat{0}$ with probability $\frac{1}{2}$, a standard Chernoff argument gives that for any constant $\varepsilon > 0$, the product distribution satisfies an $\varepsilon$-tight concentration on the relative $\bF_{2^n}$-Hamming weight of $\frac{1}{2}$, i.e.
\begin{align}
    \label{eqn:errorDistrTightness}
    \Pr_{\mat{e} \sim \prod_{i \in [m]} \cD_i } \left [\left( \frac{1}{2} - \varepsilon\right) \cdot m \leq \hw_{2^n}(\mat{e}) \leq  \left( \frac{1}{2} + \varepsilon \right) \cdot m  \right ] \geq 1 - 2^{-\Omega(\lambda)}.
\end{align}

Recalling the construction of $\decode$, the first requirement for correctness of this algorithm is that, given a noisy codeword of the form $\mat{z} = \mat{x} + \mat{e}$ where $\mat{x}$ is the original noisefree codeword in $C^\perp$, the list-decoding of $C^\perp$ succeeds in returning a list $L$ that contains $\mat{x}$.
This occurs as long as the list-decoding radius is sufficiently large enough to encompass the typical error, where the term typical is formally captured above in Equation~\ref{eqn:errorDistrTightness} on the error distribution. Using the list-decoding bound found in~\cite{FOCS:GurSud98}, this requirement is therefore,
\[m - \sqrt{(1-\alpha) m^2} > \left( \frac{1}{2} + \varepsilon \right) \cdot m,\]
which holds whenever $\alpha > \frac{3}{4} - \varepsilon + \varepsilon^2$.
This constraint will be subsumed by a similar constraint on $\alpha$ that we introduce subsequently for technical reasons. 
Observe that this constraint on $\alpha$ is quite relaxed because $\varepsilon$ can be any arbitrarily small positive constant.

The second requirement for correctness of this decoding algorithm is that among the codewords in the list $L$, there should only be one codeword $\mat{x}' \in L$ such that $\mat{z} - \mat{x}'$ satisfies the typicality condition in Equation~\ref{eqn:errorDistrTightness}. 
A sufficient condition for when this event occurs is when the error $\mat{e}$ comes from the set
\begin{align*}
    G' \coloneqq \left \{ \mat{e} \in \bF_{2^{n}}^{m} : \hw_{2^n}(\mat{e}) \leq \left(\frac{1}{2} + \varepsilon\right) m ~\bigwedge~ \forall~\mat{y}\in C^{\perp} \setminus \{0\},~\hw_{2^n}(\mat{e} - \mat{y}) >  \left(\frac{1}{2} + \varepsilon\right)m \right\}.
\end{align*}
Observe that the condition $ \hw_{2^n}(\mat{e}) \leq \left(\frac{1}{2} + \varepsilon\right) m$ implies that $\mat{z} - \mat{x} = \mat{e}$ satisfies the typicality constraint, while for all $\mat{x}' \neq \mat{x} \in L \subseteq C^\perp$, the vector $\mat{z} - \mat{x}'$ has relative $\bF_{2^n}$-Hamming weight exceeding the typicality constraint. 
Therefore, whenever a sampled $\mat{e}$ is in $G'$, then $\decode$ will successfully uniquely decode the noisy codeword $\mat{z}$.

We now argue that the probability that an error $\mat{e}$ sampled from $\prod_{i \in [m]} \cD_i $ lands in the set $G'$ is at least $1 - 2^{-\Omega(\lambda)}$.
First, we define the set of typical errors:
\begin{align*}
    T \coloneqq \left \{ \mat{e} \in \bF_{2^{n}}^{m} :  \left(\frac{1}{2} - \varepsilon\right) m \leq  \hw_{2^n}(\mat{e}) \leq \left(\frac{1}{2} + \varepsilon\right) m \right\}.
\end{align*}
Observe that Equation~\ref{eqn:errorDistrTightness} stipulates that $\Pr_{\mat{e} \sim~\prod_{i \in [m]} \cD_i} \left[ \mat{e} \notin T \right] \leq 2^{-\Omega(\lambda)}$.
Then,
\begin{align}
    & \Pr_{\mat{e} \sim~\prod_{i \in [m]} \cD_i} \left[ \mat{e} \notin G' \right] \notag \\
    & \qquad = \Pr_{\mat{e} \sim~\prod_{i \in [m]} \cD_i} \left[ \mat{e} \notin G' \land \mat{e} \in T \right] + \Pr_{\mat{e} \sim~\prod_{i \in [m]} \cD_i} \left[ \mat{e} \notin G' \land \mat{e} \notin T \right] \tag{Law of total probability.} \notag \\
    & \qquad \leq \Pr_{\mat{e} \sim~\prod_{i \in [m]} \cD_i} \left[ \mat{e} \notin G' \land \mat{e} \in T \right] + \Pr_{\mat{e} \sim~\prod_{i \in [m]} \cD_i} \left[ \mat{e} \notin T \right] \notag \\
    & \qquad \leq \Pr_{\mat{e} \sim~\prod_{i \in [m]} \cD_i} \left[ \mat{e} \notin G' \land \mat{e} \in T \right] + 2^{-\Omega(\lambda)}. \label{eqn:decodeCorrect1}
\end{align}
We now upper bound the first term in the above sum in which we have the event $\mat{e} \in \bF_{2^n}^m$ such that $ \mat{e} \notin G' \land \mat{e} \in T$. To give an upper bound, we formulate a necessary event for this event, and upper bound the probability of the necessary event, which upper bounds the probability of this event.
Observe that
\begin{enumerate}
    \item If $\mat{e} \in T$, then there exists a subset of indices $S^* \subseteq [m]$ on which $\mat{e}_i = \mat{0} \in \bF_{2^n}$ such that $\abs{S^*} \geq \left( \frac{1}{2} - \varepsilon \right) m$.
    \item If $\mat{e} \notin G'$, then there exists a codeword $\mat{y}^* \in  C^\perp \setminus \{0\}$ such that $\hw_{2^n}(\mat{e} - \mat{y}^*) \leq \left(\frac{1}{2} + \varepsilon \right) m$.
\end{enumerate}
We can rewrite the latter observation as, 
\begin{align}
    \label{eqn:rewrittenHW}
    \hw_{2^n} ( \mat{e} - \mat{y}^*)  = \hw_{2^n} (\mat{e}_{S^*} - \mat{y}^*_{S^*}) +  \hw_{2^n} (\mat{e}_{\overline{S^*}} - \mat{y}^*_{\overline{S^*}})  \leq \left(\frac{1}{2} + \varepsilon \right) m.
\end{align}
where the subscript with a set denotes a restriction to the indices in the set, i.e. $\mat{e}_{S^*} = \mat{e}_{i_1} \mat{e}_{i_2} \cdots \mat{e}_{i_{\abs{S^*}}}$ for $i_1 < i_2 < \cdots < i_{\abs{S^*}} \in S^*$, and where $\overline{S^*} = [m] \setminus S^*$. Then using the structure of the Reed-Solomon code, we have that
\begin{align}
    \hw_{2^n}(\mat{e}_{S^*} - \mat{y}^*_{S^*}) &= \hw_{2^n} (\mat{y}^*_{S^*}) \notag\\
    &\geq \abs{S^*} - (1-\alpha)m + 1 \notag\\
    &\geq \left(\frac{1}{2} - \varepsilon \right)m - (1-\alpha)m + 1.
    \label{eqn:LBOnStar}
\end{align}
The first inequality is because the codeword $\mat{y}^*$ is a vector of evaluations of a univariate polynomial of degree $(1-\alpha)m - 1$, which can can have at most $(1-\alpha)m - 1$ many roots.
The second inequality is due to the fact that $\mat{e} \in T$ as noted above.

Equation~\ref{eqn:rewrittenHW} and Equation~\ref{eqn:LBOnStar} imply the following useful necessary event for when $\mat{e}$ lands in the set $G'$: Namely we will consider, for any fixed $S^* \subseteq [m]$ such that $\left( \frac{1}{2} - \varepsilon \right) m \leq\abs{S^*} \leq \left( \frac{1}{2} + \varepsilon\right) m$, the event $\cE_{S^*}$ over $\mat{e} \sim \prod_{i \in [m]} \cD_i$ that
\begin{align}
    \label{eqn:necessaryCondition}
     \exists~\mat{y}^* \in C^{\perp} \setminus \{\mat{0}\}, \text{ s.t. } \hw_{2^n} (\mat{e}_{\overline{S^*}} - \mat{y}^*_{\overline{S^*}}) \leq \left(2\varepsilon - \alpha + 1 \right) m - 1.
\end{align}
Let $S_{\mat{e}} \coloneqq \left \{ i \in [m] : \mat{e}_i = \mat{0} \in \bF_{2^n}  \right\}$, then observe that Equation~\ref{eqn:rewrittenHW} and Equation~\ref{eqn:LBOnStar} imply that for any $S^* \subseteq [m]$ such that $\left( \frac{1}{2} - \varepsilon \right) \leq \abs{S^*} \leq \left( \frac{1}{2} + \varepsilon\right) m$
\begin{align}
    \Pr_{\mat{e} \sim \prod_{i \in [m]} \cD_i} \left[ \mat{e \notin G' \mid S_{\mat{e}} = S^*} \right] \leq \Pr_{\mat{e} \sim \prod_{i \in [m]} \cD_i} \left[ \cE_{S^*} \right] .
\end{align}
By a straightforward counting argument, we can show that for any $S^* \subseteq [m]$ such that $\left( \frac{1}{2} - \varepsilon \right) \leq \abs{S^*} \leq \left( \frac{1}{2} + \varepsilon\right) m$,  we have
\begin{align}
    & \lg \left(\Pr_{\mat{e} \sim \prod_{i \in [m]} \cD_i} \left[ \cE_{S^*} \right ]  \right) \notag \\
    & \qquad \coloneqq \lg \left( \Pr_{\mat{e} \sim \prod_{i \in [m]} \cD_i} \left [ \exists~\mat{y}^* \in C^{\perp} \setminus \{ \mat{0} \} \text{ s.t. } \hw_{2^n} \left(\mat{e}_{\overline{S^*}} - \mat{y}^*_{\overline{S^*}}  \right) \leq (2 \varepsilon - \alpha + 1)m - 1 \right] \right) \notag \\
    & \qquad \leq - 2 \left(\alpha -  \left (1 - \frac{c_{\min}}{4} + \varepsilon \left (1 - \frac{c_{\min}}{2} \right ) \right) \right) \cdot mn + O(m \log m) \label{eqn:postUnionBound}.
\end{align}
The probability inside the $\lg$ is therefore $2^{-\Omega(mn)}$ whenever $\alpha > \left (1 - \frac{c_{\min}}{4} + \varepsilon \left (1 - \frac{c_{\min}}{2} \right ) \right)$.
To prove Equation~\ref{eqn:postUnionBound}, we first show that for any $S^* \subseteq [m]$ such that $\left( \frac{1}{2} - \varepsilon \right) \leq\abs{S^*} \leq \left( \frac{1}{2} + \varepsilon\right) m$, for any fixed $\mat{y}^* \in C^{\perp} \setminus \{ \mat{0} \}$,
\begin{align}
    &\lg \left ( \Pr_{\mat{e} \sim \prod_{i \in [m]} \cD_i} \left [\hw_{2^n} \left(\mat{e}_{\overline{S^*}} - \mat{y}^*_{\overline{S^*}}  \right) \leq (2 \varepsilon - \alpha + 1)m - 1 \right] \right) \notag \\
    & \qquad \leq \left(2 \varepsilon - \alpha + 1 - c_{\min} \left ( \frac{1}{2} + \varepsilon \right) \right) mn + O(m \log m). \label{eqn:preUnionBound}
\end{align}
and then apply a union bound over all $2^{(1-\alpha) m n}$ many codewords of $C^{\perp}$.
To show Equation~\ref{eqn:preUnionBound}, for any $\mat{y}^* \in C^{\perp} \setminus \{ \mat{0} \}$, the number of $\mat{e}_{\overline{S^*}}$ that satisfy $\hw_{2^n} \left( \mat{e}_{\overline{S^*}} - \mat{y}^*_{\overline{S^*}} \right) \leq (2\varepsilon - \alpha + 1)m - 1$ is at most 
\begin{align*}
    & \sum_{i=1}^{ (2\varepsilon - \alpha + 1)m - 1} \binom{\abs{\overline{S^*}}}{i} \cdot (2^n)^{i}\\
    & \qquad \qquad  \leq  \left( (2\varepsilon - \alpha + 1)m - 1 \right) \cdot \binom{\left( \frac{1}{2} + \varepsilon \right) m}{(2\varepsilon - \alpha + 1)m - 1} \cdot 2^{n\cdot ((2\varepsilon - \alpha + 1)m - 1)}
\end{align*}
where the inequality holds when $\alpha \geq \frac{3}{2}\varepsilon + \frac{3}{4}$, subsuming the previous constraint that $\alpha > \frac{3}{4} - \varepsilon + \varepsilon^2$.
Each such $\mat{e}_{\overline{S^*}}$ is sampled with probability $2^{- \sum_{i \in \overline{S^*}} c_i n} \geq 2^{- c_{\min} \left( \frac{1}{2} + \varepsilon \right) mn }$. Combining these two facts gives Equation~\ref{eqn:preUnionBound}.
By applying a union bound over all $2^{(1-\alpha) m n}$ many codewords of $C^{\perp}$ on Equation~\ref{eqn:preUnionBound}, we obtain Equation~\ref{eqn:postUnionBound}.
Then observe that
\begingroup
\allowdisplaybreaks
\begin{align}
    & \Pr_{\mat{e} \sim~\prod_{i \in [m]} \cD_i} \left[ \mat{e} \notin G' \land \mat{e} \in T \right] \notag\\
    & \qquad = \Pr_{\mat{e} \sim~\prod_{i \in [m]} \cD_i} \left[\mat{e} \in T \right] \cdot \Pr_{\mat{e} \sim~\prod_{i \in [m]} \cD_i} \left[ \mat{e} \notin G' \mid \mat{e} \in T \right] \notag \\
    & \qquad \leq 1 \cdot \Pr_{\mat{e} \sim~\prod_{i \in [m]} \cD_i} \left[ \mat{e} \notin G' \mid \mat{e} \in T \right] \notag \\
    & \qquad = \sum_{\substack{S^* \subseteq [m]\\ \left(\frac{1}{2} - \varepsilon \right) m \leq \abs{S^*} \leq \left(\frac{1}{2} + \varepsilon \right) m}} \Pr_{\mat{e} \sim~\prod_{i \in [m]} \cD_i} \left[ S_{\mat{e}} = S^* \mid \mat{e} \in T \right] \Pr_{\mat{e} \sim~\prod_{i \in [m]} \cD_i} \left[ \mat{e} \in G' \mid S_{\mat{e}} = S^* \right] \notag \\
    & \qquad \leq \sum_{\substack{S^* \subseteq [m]\\ \left(\frac{1}{2} - \varepsilon \right) m \leq \abs{S^*} \leq \left(\frac{1}{2} + \varepsilon \right) m}} \Pr_{\mat{e} \sim~\prod_{i \in [m]} \cD_i} \left[ S_{\mat{e}} = S^* \mid \mat{e} \in T \right] \Pr_{\mat{e} \sim~\prod_{i \in [m]} \cD_i} \left[ \cE_{S^*} \right] \notag \\
    & \qquad \leq 2^{-\Omega(mn)} \cdot \sum_{\substack{S^* \subseteq [m]\\ \left(\frac{1}{2} - \varepsilon \right) m \leq \abs{S^*} \leq \left(\frac{1}{2} + \varepsilon \right) m}} \Pr_{\mat{e} \sim~\prod_{i \in [m]} \cD_i} \left[ S_{\mat{e}} = S^* \mid \mat{e} \in T \right] \notag \\
    & \qquad =  2^{-\Omega(mn)}. \label{eqn:decodeCorrect2}
\end{align}
\endgroup
Combining Equation~\ref{eqn:decodeCorrect1} and Equation~\ref{eqn:decodeCorrect2} gives
\begin{align*}
     \Pr_{\mat{e} \sim~\prod_{i \in [m]} \cD_i} \left[ \mat{e} \notin G' \right] \leq 2^{-\Omega(mn)} + 2^{-\Omega(\lambda)}.
\end{align*}
Since $\mat{e} \in G'$ is a sufficient condition for correctness of $\decode$, this completes the proof of our desired lemma.
\end{proof}

\subsubsection{Error Distribution Induced by Shift-Invariant and 2-wise Independent Distributions}

Our goal is to characterize the distribution of errors induced by any shift-invariant and 2-wise independent polynomial distribution over multivariate $\bF_2$ polynomials.
We show that the induced distribution of errors satisfies the condition in Lemma~\ref{lemma:decodableBurstErrors}, i.e. they are uniquely decodable under a Reed-Solomon code.
Namely, we will prove the following lemma.

\begin{lemma}
    \label{lemma:4wiseErrorDistr}
    Let $\lambda \in \bN$. Let $n = \Theta(\lambda)$, $m = \omega( \log (\lambda))$, and let $\{ X_{i, j} \}_{i \in [m], j \in [n]}$ be indeterminates.
    Let $\cP = (p_1, \ldots, p_m)$ be a random variable in which, for $i \in [m]$, where $p_i \sim \sP_i$ where $\sP_i$ is a 2-wise independent distribution over $\bF_2[X_{i, 1}, \ldots, X_{i, n}]$, 
    For $i \in [m]$, let $\cD_i$ denote the distribution with probability mass function 
    $\E_{p_i} \left[ \abs{\hat{W}^{p_i}(\cdot)}^2 \right]$
    and let $\cD \coloneqq \prod_{i \in [m]} \cD_i$.
    Then, for sufficiently large $\lambda \in \bN$ such that $n(\lambda) \geq 10$,
    \begin{itemize}
        \item For all $i \in [m]$, $\cD_i$ is a distribution over $\bF_2^{n}$ that takes $0$ with probability $\frac{1}{2}$ and otherwise takes any element in the support with probability at most $2^{-n/2}$.
    \end{itemize} 
\end{lemma}
\begin{proof}
    This is immediate by Lemma~\ref{lemma:MdDistr}.
\end{proof}

\begin{lemma}
    \label{lemma:MdDistr}
    For all $n \in \bN$ such that $n \geq 10$, for any $2$-wise independent distribution $\sP$ over $\bF_2[X_1, \ldots, X_n]$,
   \begin{align*}
    \E_{p\sim \sP} \left[ \abs{\hat{W}^{p}(\mat{0})}^2 \right] &= \frac{1}{2}.\\
    \forall \mat{e} \in \bF_{2^n}\setminus\{\mat{0}\}, \E_{p \sim \sP} \left[ \abs{\hat{W}^{p}(\mat{e})}^2 \right] & \leq 2^{-n/2}.
    \end{align*}
\end{lemma}

\begin{proof}
    For any polynomial $p$, let $N_p$ denote the number of roots. 
    Opening the definition of $\hat{W}^{p}$ gives 
    \begin{align*}
        \abs{\hat{W}^{p}(\mat{0})}^2 &= \left( \frac{1}{2^{n/2}} \sum_{\mat{z} \in \bF_{2^n}} W^{p}(\mat{z}) \right)^2 \\
        & = \left( \frac{1}{2^{n/2}} \cdot  \frac{N_p}{\sqrt{N_p}} \right)^2  = \frac{N_p}{2^n}.
    \end{align*}
    For any $\mat{x} \in \bF_{2}^{n}$, define the scaled indicator random variable $\Tilde{\bI}_{\mat{x}}$ as a function of $p$:
    \begin{align*}
        \Tilde{\bI}_{\mat{x}}(p) = \begin{cases} \frac{1}{2^{n}} & \text{if } p(\mat{x}) = 0\\
        0 & \text{o.w.}
        \end{cases}
    \end{align*}  and observe that for all $\mat{x} \in \bF_{2}^{n}$, we have $\E_{p} \left[ \Tilde{\bI}_{\mat{x}}(p) \right] = \frac{1}{2^{n+1}}$.
    Then, 
    \begingroup
    \allowdisplaybreaks
    \begin{align*}
        \E_{p} \left[ \abs{\hat{W}^{p}(\mat{0})}^2 \right] &= \E_{p} \left[ \frac{N_p}{2^n} \right]\\
        &= \E_{p} \left[ \sum_{\mat{x} \in \bF_{2}^{n}} \Tilde{\bI}_{\mat{x}}(p) \right]\\
        &= \sum_{\mat{x} \in \bF_{2}^{n}} \E_{p} \left[ \Tilde{\bI}_{\mat{x}}(p) \right] = \frac{1}{2}.
    \end{align*}
    \endgroup

    \noindent Next we will prove that $$ \forall \mat{e} \in \bF_{2^n}\setminus\{\mat{0}\}, \E_{p} \left[ \abs{\hat{W}^{p}(\mat{e})}^2 \right]  \leq 2^{-\Omega(n)}.$$
    We abuse notation and a vector $\mat{z} \in \bF_{2^n}$ will be occasionally considered in the isomorphic vector space  $\mat{z} = (z_1, \ldots, z_n) \in \bF_{2}^{n}$.
    Expand the definition of  $\hat{W}^{p}(\mat{e})$, 
    \[ \hat{W}^{p}(\mat{e})=\frac{1}{2^{n/2}} \sum_{\mat{z} \in \extension} W^{p}(\mat{z}) (-1)^{\trace_{2^{n}}(\mat{e} \cdot \mat{z})}. \]
    Let $\sigma : \bF_{2^n} \to \bF_2^{n}$ be the bijective map whose existence is guaranteed by Lemma~\ref{lemma:traceFieldExt}.
    Then,
    \[ \hat{W}^{p}(\mat{e})=\frac{1}{2^{n/2}} \sum_{\mat{z} \in \bF_{2^n}} W^{p}(\mat{z}) (-1)^{\ipt{\sigma(\mat{e})}{\mat{z}}} \]
    where we interpret $\mat{z}$ in  the exponent $\ipt{\sigma(\mat{e})}{\mat{z}}$ as a $\bF_{2}^n$ vector via the canonical isomorphism between $\bF_{2^n}$ and $\bF_2^{n}$. 
Observe for all $\mat{z} \in \bF_{2^n}$ by definition of $W^{p}(\mat{z})$,
\[W^{p}(\mat{z}) = \frac{1}{2 \sqrt{N_p}} \cdot \left(1 + (-1)^{p(\mat{z})} \right).\]
By substitution,
\begin{align*}
    \hat{W}^{p}(\mat{e})&=\frac{1}{2^{n/2 +1} \sqrt{N_p}} \cdot \sum_{\mat{z} \in \bF_{2}^{n}} \left( 1 + (-1)^{p(\mat{z})} \right) \cdot (-1)^{\ipt{\sigma(\mat{e})}{\mat{z}}}\\
    &= \frac{1}{2^{n/2 +1} \sqrt{N_p}} \cdot \left(  \sum_{\mat{z} \in \bF_{2}^{n}}  (-1)^{\ipt{\sigma(\mat{e})}{\mat{z}}}  + \sum_{\mat{z} \in \bF_{2}^{n}}  (-1)^{p(\mat{z}) + \ipt{\sigma(\mat{e})}{\mat{z}}}   \right)\\
    &=  \frac{1}{2^{n/2 +1} \sqrt{N_p}}  \cdot \sum_{\mat{z} \in \bF_{2}^{n}}  (-1)^{p(\mat{z}) + \ipt{\sigma(\mat{e})}{\mat{z}}}
\end{align*}
where in the last equality we critically  used the fact that $\mat{e} \neq \mat{0} \in \bF_{2^n}$ implies that $\sigma(\mat{e}) \neq \mat{0}$ by Lemma~\ref{lemma:traceFieldExt}, implying that
\[ \sum_{\mat{z} \in \bF_{2}^{n}}  (-1)^{\ipt{\sigma(\mat{e})}{\mat{z}}}  = 0.\]
First let us define $P(\mat{z})=(-1)^{p(\mat{z})}$. Observe that $P:\{0,1\}^n \rightarrow \{+1,-1\}$.
Therefore we can rewrite for $\mat{e} \neq \mat{0}\in \bF_{2^n}$,
\begin{align*}
    \hat{W}^{p}(\mat{e})&=\frac{1}{2^{n/2 +1} \sqrt{N_p}} \cdot \frac{2^{n/2}}{2^{n/2}} \cdot \sum_{\mat{z} \in \bF_{2}^{n}} P(\mat{z}) \cdot (-1)^{\ipt{\sigma(\mat{e})}{\mat{z}}}\\
    &=\frac{2^{n/2}}{2\sqrt{N_p}} \cdot \frac{1}{2^{n}} \cdot \sum_{\mat{z} \in \bF_{2}^{n}} P(\mat{z}) \cdot (-1)^{\ipt{\sigma(\mat{e})}{\mat{z}}}\\
    &=\frac{2^{n/2}}{2\sqrt{N_p}} \cdot \E_{\mat{z} \in \bF_2^n} \left [(-1)^{\ipt{\sigma(\mat{e})}{\mat{z}}} P(\mat{z}) \right]=\frac{2^{n/2}}{2\sqrt{N_p}} \cdot \tilde{P}(\sigma(\mat{e})),
\end{align*}
where we define the function $\tilde{P}$ such that for any $\mat{e} \in \bF_{2}^{n}$, \[\tilde{P}(\mat{e})\coloneqq \E_{\mat{z} \in \bF_2^n}\left [(-1)^{\ipt{\mat{e}}{\mat{z}}} \cdot P(\mat{z}) \right]=\frac{1}{2^{n}} \cdot \sum_{\mat{z} \in \extension} P(\mat{z}) \cdot (-1)^{\ipt{\mat{e}}{\mat{z}}}.\]

\noindent
We begin by splitting the support of $\sP$ into two subsets $\mathsf{Central} \sqcup \mathsf{Tail}$ where
\begin{align*}
\mathsf{Central} & \coloneqq \left \{p \in \supp (\sP) : N_p \in [2^{n-1}-2^{0.9n - \offsetC},2^{n-1}+2^{0.9n - \offsetC}] \right\}\\
\mathsf{Tail} & \coloneqq \supp (\sP) \setminus \mathsf{Central} 
\end{align*}

\noindent
Define $\delta \coloneqq \Pr_p \left[ p \in \mathsf{Tail} \right]$ and observe by a standard Chebyshev's inequality
\begin{align*}
    \delta &= \Pr_{p} \left[ \abs{N_p - 2^{n-1}} > 2^{0.9 n - \offsetC} \right ]\\
    &\leq \frac{\E_{p} \left[(N_p - 2^{n-1})^2 \right]}{2^{1.8 n - \twoOffsetC}} \\
    & = \frac{2^{n-2}}{2^{1.8 n - \twoOffsetC }} = 2^{-0.8n}.~\tag{By Lemma~\ref{lemma:moments}.} 
\end{align*}
\noindent
Recall that for all $p \in \bF_2[X_1, \ldots, X_n]$ we have $\sum_{\mat{e} \in \bF_{2^n}} \abs{\hat{W}^{p}(\mat{e})}^2 = \sum_{\mat{e} \in \bF_{2^n}} \abs{W^{p}(\mat{e})}^2 = N_p / N_p = 1$ where the first equality is due to Parseval's Theorem (Lemma~\ref{lem:parseval}). This implies that for all $\mat{e} \in \bF_{2^n}$, for all $p \in \bF_2[X_1, \ldots, X_n]$, we have $\abs{\hat{W}^{p}(\mat{e})}^2 \leq 1$.
Therefore, for any $\mat{e} \neq \mat{0}$,
\begin{align*}
    \E_p \left[\abs{\hat{W}^p(\mat{e})}^2 ~\bigg\vert~  p \in \mathsf{Tail} \right] \leq 1.
\end{align*}
Then observe that for $\mat{e} \neq \mat{0} \in \bF_2^n$ and for $n \geq 10$, 
\begingroup
\allowdisplaybreaks
\begin{align*}
    \E_p \left [\abs{ \hat{W}^p(\mat{e})}^2 ~\bigg\vert~ p \in \mathsf{Central} \right] &= \E_p \left [\frac{2^{n-2}}{N_p} \cdot \tilde{P}(\mat{e})^2 ~\bigg\vert~  p \in \mathsf{Central} \right]\\
    & \leq \frac{2^{n-2}}{2^{n-1} - 2^{0.9 n - \offsetC}} \cdot \E_p \left [\tilde{P}(\mat{e})^2 ~\bigg\vert~  p \in \mathsf{Central} \right]\\
    &= \frac{2^{n-2}}{2^{n-1} - 2^{0.9 n - \offsetC}} \cdot \left(  \frac{1}{1- \delta} \cdot \E_p \left[\tilde{P}(\mat{e})^2 \right] - \frac{\delta}{1 - \delta}  \cdot \E_p \left[\tilde{P}(\mat{e})^2 ~\big\vert~ p \in \mathsf{Tail} \right] \right) \\
    & \leq \frac{2^{n-2}}{2^{n-1} - 2^{0.9 n - \offsetC}} \cdot \frac{1}{1- \delta} \cdot \E_p \left[\tilde{P}(\mat{e})^2 \right]  \\
    & \leq 1 \cdot 2 \cdot \frac{1}{2^{n}} = 2^{- n + 1},
\end{align*}
\endgroup
where the first inequality uses the conditioning on $p \in \mathsf{Central}$ to imply that $N_p \geq 2^{n-1} - 2^{0.9n - \offsetC}$, the next line (third line) uses the total law of probability, the fourth line uses the non-negativity of $\tilde{P}(\mat{e})^2$, and the fifth line uses Lemma~\ref{lem:non0equality}.
Finally, for $n \geq 10$,
\begin{align*}
    &\E_p  \left [\abs{\hat{W}^p(\mat{e})}^2 \right]\\
    & \qquad =\Pr_p \left [p \in \mathsf{Tail} \right ] \cdot \E_p \left[\abs{\hat{W}^p(\mat{e})}^2 ~\bigg\vert~  p \in \mathsf{Tail} \right]+\Pr_p \left [p \in \mathsf{Central} \right] \cdot \E_p \left [\abs{ \hat{W}^p(\mat{e})}^2 ~\bigg\vert~ p \in \mathsf{Central} \right]\\
    & \qquad \leq 2^{-0.8 n} \cdot 1 +  1 \cdot 2^{-n + 1} \leq 2^{-n/2}.
\end{align*}

\end{proof}

\noindent
In the above lemma, we used the following fact.

\begin{lemma}
    \label{lem:non0equality}
Let $P : \{0, 1\}^n \to \{ \pm 1\}$ and $\tilde{P} : \{0, 1\}^n \to \bR$ be defined as above relative to a polynomial $p \in \bF_2[X_1, \ldots, X_n]$.
For all $\mat{e} \in \bF_{2}^n$, for any 2-wise independent distribution $\sP$ over $\bF_2[X_1, \ldots, X_n]$,  \[\E_{p \sim \sP}[\tilde{P}(\mat{e})^2]=\frac{1}{2^n}.\]
\end{lemma}
\begin{proof}
    \begin{align*}
        \E_p[\tilde{P}(\mat{e})^2] & = \E_p\left[\left(\frac{1}{2^{n}} \cdot \sum_{\mat{z} \in \bF_2^{n}} P(\mat{z}) \cdot (-1)^{\ipt{\mat{e}}{\mat{z}}}\right)^2\right]\\
         & = \E_p\left[\frac{1}{2^{2n}} \cdot \sum_{\mat{z}, \mat{z}' \in \bF_2^{n}} P(\mat{z})\cdot P(\mat{z}') \cdot (-1)^{\ipt{\mat{e}}{\mat{z} + \mat{z}'}} \right]\\
        &= \frac{1}{2^{2n}}  \sum_{\mat{z}, \mat{z}' \in \bF_2^{n}} \E_p\left[ P(\mat{z})\cdot P(\mat{z}') \right] \cdot (-1)^{\ipt{\mat{e}}{\mat{z} + \mat{z}'}}\\
        &= \frac{1}{2^{2n}} \left [ \sum_{\mat{z} \in \bF_2^{n}}  \E_p\left[ P(\mat{z})^2 \right] +  \sum_{\mat{z}\neq \mat{z}' \in \bF_2^{n}} \E_p\left[ P(\mat{z})\cdot P(\mat{z}') \right] \cdot (-1)^{\ipt{\mat{e}}{\mat{z} + \mat{z}'}}  \right]\\
        &= \frac{1}{2^{n}}
    \end{align*}
    Where in the last equality, 2-wise independence directly implies that for $\mat{z} \neq \mat{z}' \in \bF_2^{n}$, 
    \[\E_p\left[ P(\mat{z})\cdot P(\mat{z}') \right] = \E_p[P(\mat{z}) ] \cdot \E_p[P(\mat{z}')] = 0,\]
    and where we use the fact that $\E_p \left[ P(\mat{z})^2 \right] = 1$ since the codomain of $P$ is $\{ \pm 1\}$.

\end{proof}

\begin{lemma}[Moments of the Variety]
    \label{lemma:moments}
    For any polynomial $p$, let $N_p$ denote the number of roots of $p$.
    For any $n \in \bN$, for any $2$-wise independent distribution $\sP$ over $\bF_2[X_1, \ldots, X_n]$,
    \begin{align*}
     \mu \coloneqq & \E_{p}[N_p] = 2^{n-1} \\
        &\E_{p}[N_p^2] = 2^{2n-2} - 2^{n-2}\\
    \end{align*}
    with the second central moment given by 
    \begin{align*}
       \E_{p} [(N_p - \mu)^2] &= 2^{n-2}.
    \end{align*}
    
\end{lemma}
\begin{proof}
    For any $\mat{x} \in \bF_2^n$, define the $0/1$ indicator random variable $\bI_{\mat{x}} : \bF_2[X_1, \ldots, X_n] \to \{0, 1\}$ such that $\bI_{\mat{x}}(p) = \begin{cases}
        1 & \text{if } p(\mat{x}) = 0\\
        0 & \text{o.w.}
    \end{cases}.$
    Then the random variable for the number of roots $N_p = \sum_{\mat{x} \in \bF_2^n} \bI_{\mat{x}}(p)$.
    Observe that for all $\mat{x} \in \bF_2^n$ we have $\bI_{\mat{x}}^2 = \bI_{\mat{x}}$.
    Observe that for all $\mat{x} \in \bF_2^n$, we have $\E_{p} \left[ \bI_{\mat{x}} \right] = \frac{1}{2}$ because of $1$-wise independence.
    By the linearity of expectation, we obtain $\mu = 2^{n-1}$.

    Observe that the 2-wise independence of $\sP$ implies that for $\mat{x} \neq \mat{y}$, we have 
    \[\E_{p} \left[ \bI_{\mat{x}} (p) \cdot \bI_{\mat{y}}(p) \right ] = \E_{p} \left[ \bI_{\mat{x}} (p)  \right ] \cdot \E_{p} \left[  \bI_{\mat{y}} (p) \right ].\]
    Therefore,
    \begingroup
    \allowdisplaybreaks
    \begin{align*}
        \E_{p}[N_p^2] &= \sum_{\mat{x} \neq  \mat{y} \in \bF_2^n} \E_{p}[\bI_{\mat{x}}(p) \cdot \bI_{\mat{y}}(p)] + \sum_{\mat{x} \in \bF_2^{n}} \E_{p}[\bI_{\mat{x}}(p)]\\
        &= \sum_{\mat{x} \neq \mat{y}\in \bF_2^{n}} \E_{p} [ \bI_{\mat{x}}(p)] \cdot \E_{p} [ \bI_{\mat{y}}(p)] + 2^{n-1}\\
        &= 2^n (2^n -1) \cdot \frac{1}{4} + 2^{n-1}\\
        &= 2^{2n-2} + 2^{n-2}
    \end{align*}
    \endgroup
\end{proof}

\subsubsection{Putting the Lemmas Together}

Finally, the proof of the main technical lemma (Lemma~\ref{lemma:mainTechnicalLemma}) combines  Lemma~\ref{lemma:decodableBurstErrors} with Lemma~\ref{lemma:4wiseErrorDistr}.

\begin{proof}[Proof of Lemma~\ref{lemma:mainTechnicalLemma}]
\label{proof:mainTechnicalLemma}

    We first prove that $\sum_{(\mathbf{x}, \mbf{e}) \in \mathsf{BAD}} \abs{\hat{V}(\mbf{x}) \hat{W}^{\cP}( \mat{e}) }^2  = \sum_{\mat{e} \in \cB} \abs{\hat{W}^{\cP}(\mat{e})}^2$ by straightforwardly opening the definitions of $\mathsf{BAD}$ and $\hat{V}$.
    \begin{align*}
        & \sum_{(\mathbf{x}, \mbf{e}) \in \mathsf{BAD}} \abs{\hat{V}(\mbf{x}) \hat{W}^{\cP}( \mat{e}) }^2 \\
        & \qquad = \sum_{(\mat{x}, \mat{e}) \in (\mextension \setminus C^{\perp}) \times \mextension} \abs{\hat{V}(\mat{x}) \hat{W}^{\cP} (\mat{e})}^2 + \sum_{(\mat{x}, \mat{e}) \in C^{\perp} \times \cB} \abs{\hat{V}(\mat{x}) \hat{W}^{\cP}(\mat{e})}^2 \tag{By definition of $\mathsf{BAD}$.}\\
        & \qquad = 0 + \frac{1}{\abs{C^{\perp}}} \cdot \sum_{(\mat{x}, \mat{e}) \in C^{\perp} \times \cB} \abs{\hat{W}^{\cP} (\mat{e})}^2  \tag{By definition of $\hat{V}$.}\\
        & \qquad = \sum_{\mat{e} \in \cB} \abs{\hat{W}^{\cP}(\mat{e})}^2.
    \end{align*}
    To argue the existence of a negligible function $\mu : \bN \to [0, 1]$ such that $\sum_{\mat{e} \in \cB} \abs{\hat{W}^{\cP}(\mat{e})}^2 \leq \mu(\lambda)$, by Markov's inequality, it suffices to argue that there exists such a negligible function $\mu$ such that 
    \begin{align}
        \E_{\cP} \left[ \sum_{\mat{e} \in \cB} \abs{\hat{W}^{\cP}(\mat{e})}^2 \right]  = \sum_{\mat{e} \in \cB}  \E_{\cP} \left[\abs{\hat{W}^{\cP}(\mat{e})}^2 \right]  \leq \mu(\lambda). \label{eqn:degreeDFinal}
    \end{align}
    This is implied by combining Lemma~\ref{lemma:decodableBurstErrors} with Lemma~\ref{lemma:4wiseErrorDistr}.

    In more detail, recall that by Lemma~\ref{lemma:decodableBurstErrors} we have
    \begin{align*}
        \Pr_{\mat{e} = \mat{e}_1\cdots \mat{e}_m~\sim~\prod_{i \in [m]} \cD_i} \left[ \forall~\mat{x} \in C^{\perp}, \decode(\mat{x} + \mat{e}) = \mat{x} \right] \geq 1 - 2^{-\Omega(\lambda)}
    \end{align*}
 for any error distribution $\cD=\prod_i \cD_i$ that satisfies the condition on the distribution in  Lemma~\ref{lemma:decodableBurstErrors}. Also recall that the definition of the set $\cB$: \[\cB \coloneqq \{\mat{e} \in \mextension\colon \exists \mat{x} \in C^\perp, \decode(\mat{x}+\mat{e})\neq x\}.\] Thus, Lemma~\ref{lemma:decodableBurstErrors} says that as long as the error distribution $\cD$ satisfies the necessary conditions, there exists a negligible function $\mu$ such that \[ \Pr_{\mat{e} = \mat{e}_1\cdots \mat{e}_m~\sim~\prod_{i \in [m]} \cD_i} \left[ \mat{e} \in \cB\right] \leq \mu(\lambda) \implies \sum_{\mat{e} \in \cB} \cD(\mat{e}) \leq \mu(\lambda).\]
   It therefore suffices to prove that the error distribution in our case indeed satisfies the conditions of Lemma~\ref{lemma:decodableBurstErrors}. This is exactly what Lemma~\ref{lemma:4wiseErrorDistr} shows.
    Specifically, the proof of Lemma~\ref{lemma:4wiseErrorDistr} implies that the error distribution defined by $\E_{\cP} \left[\abs{\hat{W}^{\cP}(\cdot)}^2 \right]$ satisfies this condition for a value of $c_{\min} \geq \frac{1}{2}$.
    Therefore, for any $1 > \alpha > \max \left (\frac{3}{4} + \frac{3}{2} \varepsilon, \frac{7}{8} + \frac{3}{4}\varepsilon \right)$, we can apply Lemma~\ref{lemma:decodableBurstErrors} to guarantee the existence of this negligible function $\mu$.   
    We can further simplify the condition on $\alpha$ by observing that when $\varepsilon \geq \frac{1}{6}$ the first term of the maximum operator implies that we need $\alpha \geq 1$ implying that $\varepsilon \geq \frac{1}{6}$ does not result in any feasible values of $\alpha$.
    When $\varepsilon < \frac{1}{6}$ however, then the second term of the maximum operator $\frac{7}{8} + \frac{3}{4}\varepsilon$ dominates.
    Therefore we require that $\varepsilon < \frac{1}{6}$ and $1 > \alpha > \frac{7}{8} + \frac{3}{4}\varepsilon$.
\end{proof}

\subsection{Uniform Random Degree Bounded Polynomials are Shift-invariant and 2-wise Independent}
\label{sec:constantDegSI4WI}

This subsection straightforwardly shows that uniform random degree bounded polynomials are shift-invariant and $2$-wise independent. Thus, this section proves the correctness of Algorithm~\ref{algo:1} when $\cP$ is a system of uniform degree $d$ multivariate polynomials, for arbitrary constant $d \geq 2$.

\begin{lemma}
    For any $n \in \bN$, for any constant $d \in \bN$, the distribution $\cP$ over $\bF_2[X_1, \ldots, X_n]$ of uniform random at most degree $d$ polynomials is shift-invariant (Def.~\ref{def:Invariant}).
\end{lemma}
\begin{proof}
    For any $\mat{z} \in \bF_2^{n}$, the map $\Pi_{\mat{z}} : \bF_2[X_1, \ldots, X_n] \to \bF_2[X_1, \ldots, X_n]$ that maps $p \in \bF_2[X_1, \ldots, X_n]$ to $p' \in \bF_2[X_1, \ldots, X_n]$ such that for all $\mat{x} \in \bF_2^{n}$,~$p'(\mat{x}) = p(\mat{x} + \mat{z})$, is a permutation. In any characteristic two field, $\Pi_{\mat{z}}(\Pi_{\mat{z}}(p)) = p$. 
    

\end{proof}


\begin{lemma}[$2$-wise Independence of Non-linear Polynomials]
    \label{lemma:4wiseIndep}
    Let $d \geq 2$ be any integer. For any $n \in \bN$, for any two distinct vectors $\mat{x} \neq \mat{y} \in \bF_2^{n}$, for any $a_1 \in \bF_2$ and $a_2 \in \bF_2$ over the choice of a uniform random inhomogeneous degree $d$ polynomial $p \in \bF_2[X_1,\ldots,X_n]$,
    \[\Pr_{p} \left[ p(\mat{x}) = a_1 \land p(\mat{y}) = a_2 \right ] = \frac{1}{4}.\]
\end{lemma}
\begin{proof}
    Observe that $\Pr_{p} \left[ p(\mat{x}) = a_1 \right ] =1/2$. 
    Then observe that the random polynomial $p(\mat{x})$ and the random polynomial $p(\mat{y})$ are independent uniform random variables since
    there exists an index $i \in [n]$ such that $\mat{x}_i \neq \mat{y}_i \in \bF_2$. 
    These two facts imply the desired probability statement. 
\end{proof}

\section{Conjectured Classical Hardness}\label{sec:classical}

Although the proof of quantum easiness is independent of the degree $d$, we instantiate our construction only for $d \geq 3$. We make this choice because, in our regime, the system instantiated with quadratic polynomials can be efficiently solved (under a very plausible assumption). Indeed, one can exploit the reduction of quadratic forms to produce extra linear equations in the system by fixing relatively few variables. We describe this approach in Section \ref{sec:quads_easy}, where we note that extending it to polynomials of degree $d \geq 3$ is unlikely to be feasible. 

The classical solving algorithms we studied all have exponential complexity in $n^2$. In Section \ref{sec:combi}, we examine algorithms that can be viewed as enhanced versions of exhaustive search. The most efficient of them has a cost of $\widetilde{\mathcal{O}}(2^{(1-\alpha)n^2})$, where the $\widetilde{\mathcal{O}}(\cdot)$ notation suppresses polynomial factors. This exponent can be improved by using more efficient enumerative algorithms for solving polynomial systems (such as~\cite{din1}), but this improvement is limited and the complexity remains exponential in $n^2$. In Section \ref{sec:generic}, we consider algebraic algorithms related to Gröbner bases~\cite{BUC65,lazard,f4,f5}. Their behavior is provably exponential for generic degree-$d$ systems where the number of equations is roughly equal to the number of variables~\cite{bardet_these,BFS15}, but their cost remains uncertain for specific structured systems that do not fall into well-studied categories (see, for example,~\cite{loci,bilin}). This partly explains why it is challenging to analyze Gröbner basis solvers as precisely as those in Section~\ref{sec:combi}. Still, we provide both experimental and theoretical evidence suggesting that they should not threaten our construction.

Finally, note that all the solving algorithms described so far do not take advantage of the efficient decoding procedure of the code $C$. However, this procedure is a key component of our quantum algorithm, raising the question of whether classical algorithms could also exploit it. Despite significant efforts, we failed to come up with potential out-of-the-box solvers that can leverage it, but we think it is an interesting direction for research. An informal discussion on this can be found in Appendix \ref{app:exploiting_C}.

\begin{remark}
In this section, and with a slight abuse of notation, we denote by $C$ a binary image of the original dual Reed-Solomon code of length $m$ and dimension $\alpha m$ over $\ff{2^n}$ that is obtained through basis extension by using a fixed basis of the $\ff{2}$-vector space $\ff{2^n}$. This code is an $\ff{2}$-linear code of length $mn$ and dimension $\alpha m n$. Finally, let us recall that $\alpha \in (\frac78,1)$ in our setting.
\end{remark}


\subsection{Specialization approach for quadratics and higher degrees}\label{sec:quads_easy}

\begin{theorem}\label{theo:quads_easy}
Assume that the system $\mathcal{P}$ in Section~\ref{sec:mqsystem} consists of quadratic equations $p_i \in \ff{q}[x_{i,1},\dots,x_{i,n}]$. Then, under an assumption that a certain system of linear equations derived from $\mathcal{P}$ has a solution with high probability, there exists a classical algorithm that solves $\mathcal{F}$ in polynomial time with high probability.
\end{theorem}
While the assumption of Theorem \ref{theo:quads_easy} is non-trivial to prove, it becomes easy to prove if the Reed-Solomon code were replaced by a random linear code of the same dimension (in which case, the system would have exponentially many solutions with very high probability). The algorithm is described in the proof of the theorem. The key observation is that every quadratic form can be efficiently written as a canonical normal form. Moreover, in our setting, the equations $p_i$ involve distinct variable blocks $\mat{x}_i$, so that their normal forms will also be variable-disjoint.
\begin{proposition}[Normal Form in Characteristic $2$~\cite{lidl1997finite}]
\label{thm:MQNF}
    Let $p \in \bF_2[x_1, \ldots, x_n]$ be a quadratic form defined as $p(\mat{x}) = \mat{x}^{\top} \mat{M} \mat{x}$, where $\mat{M} \in \bF_2^{n \times n}$ is upper triangular. Let $r$ be the rank of the skew-symmetric matrix $\mat{M}+\mat{M}^{\perp}$, which is an even number. Then $p$ is equivalent to one of the following types of quadratic forms:
    \begin{enumerate}
        \item $x_1x_2 + x_3x_4 + \cdots + x_{r-1} x_{r}$.
        \item $x_1x_2 + x_3 x_4 + \cdots + x_{r-1}x_{r} + x_{r-1}^2 + x_{r}^2$.
    \end{enumerate}
\end{proposition}

\begin{proof}[Proof of Theorem~\ref{theo:quads_easy}]
    The algorithm starts by applying an invertible linear change of variables $\mat{x}_i \mapsto \mat{y}_i = (y_{i,1},\dots,y_{i,n})$ in each equation $p_i(\mat{x}_i) \in \mathcal{P}$, yielding a new equation $p'_i(\mat{y}_i) = q_i(\mat{y}_i) + \ell_i(\mat{y}_i)$, where $q_i$ is the normal form of the quadratic part of $p_i$ and $\deg{(\ell_i)} \leq 1$. One can without loss of generality assume that $q_i$ is of the form of Case 1 in Proposition \ref{thm:MQNF} by merging potential squares into $\ell_i$. Unless additional conditions are imposed on the coefficients, note that this procedure does not apply to general MQ systems in which all equations share the same set of variables. 

The rest of the algorithm exploits the sparsity of the $q_i$'s. Without loss of generality, let us assume that $n$ is even and that the rank of $q_i$ is maximal for $i \in [m]$. For each $i \in [m]$, the algorithm arbitrarily fixes the $\frac{n}{2}$ variables $y_{i,j}$ with even indices $j$ in $p'_i$. The structure of $q_i$ ensures that, after specialization, the resulting equation in the remaining variables $y_{i,j}$ with odd indices $j$ is of degree at most one. As a result, this procedure yields an affine linear system of $m$ equations in the $\frac{mn}{2}$ variables $y_{i,j}$, $i \in [m]$ and odd $j \in [n]$. 

In order to solve $\mathcal{F}$, the algorithm finally combines this linear system with the $(1-\alpha)mn$ parity check equations of $C$, applying the same change of variables and the same specialization to them before the merge. This gives an affine linear system of $m + (1-\alpha)mn$ equations in the same $\frac{mn}{2}$ variables $y_{i,j}$. This system is largely underdetermined for $\alpha > \frac{1}{2}$. In case solutions exist (which would occur with very high probability in case the Reed-Solomon code is replaced by a random linear code of the same dimension), an arbitrary solution can be found in polynomial time and easily extended to a solution of $\mathcal{F}$.
\end{proof}

\paragraph{Generalization to higher-degree polynomials.} The above algorithm highly relies on the fact that each equation of $\mathcal{P}$ involves distinct variables, which remains true in our construction where $d \geq 3$. However, we show that even a generalized form of this algorithm fails when $d=3$, and similarly for higher degrees.

Before that, recall that the algorithm in the quadratic case selected a special set of $\frac{n}{2}$ linear variable expressions (corresponding to $\frac{n}{2}$ variables after a linear change of variables), such that assigning \emph{any} $\frac{n}{2}$ Boolean values of these expressions converts each quadratic polynomial into an affine linear polynomial. This strategy can be related to techniques commonly employed for undetermined MQ systems \cite{thomae2021, furue_pqc, hashimoto2023}. The key difference is that, for random systems, these algorithms select \emph{specific} Boolean linear expressions to cancel out the quadratic terms, resulting in affine linear polynomials. Note also that no efficient generalization of the techniques in \cite{thomae2021, furue_pqc, hashimoto2023} to higher-degree systems is known. In the following, we show that such a generalization also fails for our system with $d=3$.


Given a polynomial $p \in \mathbb{F}_2[x_1,\ldots,x_n]$ and an affine linear polynomial $\ell = \alpha_0 + \sum_{j=1}^n \alpha_j x_j \in \mathbb{F}_2[x_1,\ldots,x_n]$ such that $\alpha_i=1$ for some index $i$, eliminating the variable $x_i$ from $p$ amounts to replacing this variable with $\alpha_0 + \sum_{j \in [n] \setminus \{i\}}^{n}\alpha_j x_j$ in the algebraic normal form of $p$ (as a sum of monomials). The degree of this result may (or may not) be smaller than the degree of $p$. Importantly, the degree of $p$ after variable elimination is independent of which variable was eliminated using $\ell$, since an invertible affine change of variables relates two polynomials obtained by eliminating different variables. More generally, given a polynomial $p \in \mathbb{F}_2[x_1,\ldots,x_n]$ and $t$ (consistent) affine linear polynomials $\ell_i \in \mathbb{F}_2[x_1,\ldots,x_n]$ for $i \in [t]$, we can derive the polynomial $p$ restricted to the corresponding affine subspace by performing the elimination process iteratively. Its degree is well-defined regardless of which variables were eliminated. 

\begin{lemma}\label{lemma:three_to_two}
Let $p \in \mathbb{F}_2[x_1,\ldots,x_n]$ be a uniformly chosen polynomial of degree 3.
Then, the probability that
there exists an affine subspace $A \subset \mathbb{F}_2^n$ of co-dimension at most $t = n - n^{3/4} = n - o(n)$ for which $p$ restricted to $A$
has degree at most 2 is at most $2^{-\Omega(n^2)}$.
\end{lemma}

\begin{proof}
Let $p \in \mathbb{F}_2[x_1,\ldots,x_n]$ be a uniformly chosen polynomial of degree $3$, and fix an affine subpace $A$ of co-dimension at most $t' \leq t$ together with $t'$ affine equations $\ell_i \in \mathbb{F}_2[x_1,\ldots,x_n]$ that span this subspace. Eliminating $t'$ variables from $p$ using $\ell_1,\ldots,\ell_{t'}$, we remain with a uniform polynomial in $n-t'$ variables.
Such a polynomial has $\binom{n-t'}{3}$ degree-3 coefficients, and is quadratic only if all of them are 0, which occurs with probability $2^{-\binom{n-t'}{3}} \leq 2^{-\binom{n-t}{3}}$. 

On the other hand, the number of possible choices for affine subspaces of dimension at most $t$ (spanned by at most $t$ affine equations, each with at most $n+1$ coefficients) is at most $2^{(n+1)t}$.

Taking a union bound over all affine subspaces of co-dimension at most $t$, we bound 
the probability that
there exists an affine subspace $A \subset \mathbb{F}_2^n$ of dimension at most $t$ for which $p$ restricted to $A$
has degree at most 2 by 
$$
2^{-\binom{n-t}{3}} 2^{(n+1)t} < 2^{-\binom{n-t}{3} + n(n+1)}. 
$$
If $t \leq n - n^{3/4}$, the probability is bounded by 
$$
2^{-\binom{n^{3/4}}{3} + n(n+1)} \leq 2^{-\Omega(n^2)}.
$$
\end{proof}

Based on Lemma \ref{lemma:three_to_two}, it remains to show that even specializing $t > n - o(n)$ (linearly transformed) variables in each $p_i$ to produce equations of degree $<$ 3 does not help to solve $\mathcal{F}$. The most favorable case would be to generate in this way $m$ affine linear equations. However, these equations combined with the parity check equations would form an affine linear system of $m + (1 - \alpha )mn$ equations in $mn - m(n - o(n)) = o(mn)$ variables. Recalling that the code rate $\alpha$ is a constant, this system will have no solution unless one is extremely lucky in choosing affine subspaces that are consistent with a solution to the system. 


\subsection{Combinatorial algorithms}\label{sec:combi}

We describe algorithms akin to exhaustive search, which we refer to as ``combinatorial''. The main idea of these algorithms is very simple. Recall our system consists of $n^3$ variables, $n^2$ equations of degree $d > 2$ over disjoint variable sets (each equation with $n$ variables), and $(1- \alpha) n^3$ linear equations induced by the Reed-Solomon code. This system is largely underdetermined; specifically, we have about $n^3 - n^2 - (1- \alpha) n^3 \approx \alpha n^3$ more variables than equations. Thus, we specialize about $\alpha n^3$ variables that make up $\alpha n^2$ degree-$d$ equations in a way that these equations are all satisfied. We remain with about 
$(1- \alpha) n^2$ degree-$d$ equations. We then eliminate the $(1- \alpha) n^3$ variables from the linear equation system using Gaussian elimination. Overall, we remain with a system of about $(1- \alpha)n^2$ degree-$d$ equations and variables. Assuming that a solution exists, it can be found by exhaustive search in time of about $2^{(1 - \alpha)n^2}$. The assumption that a solution exists with high probability is non-trivial, but plausible (and easy to prove if the Reed-Solomon code were replaced with a random linear code of the same dimension).

Theorem~\ref{theorem:best_combinatorial} gives a precise algorithm that achieves the claimed complexity of $2^{(1 - \alpha)n^2}$. This algorithm simply consists of sampling solutions to $\mathcal{P}$ and verifying whether they belong to the code $C$, and its complexity relies on an assumption that we now describe. 

The dual Reed-Solomon code admits a systematic generator matrix $\mat{G} = \begin{bmatrix}
\mat{I}_{\alpha m} & \mat{P}
\end{bmatrix} \in \ff{2^n}^{\alpha m  \times m },~\mat{P} \in \ff{2^n}^{\alpha m \times (1-\alpha)m}$, where $\rk{(\mat{P})} = \min{(\alpha m ,(1-\alpha)m)}$ since Reed-Solomon codes are MDS. By choosing the extension field basis appropriately, we can further make sure that the binary image $C$ has a generator matrix $\overline{\mat{G}} = \begin{bmatrix}
\mat{I}_{\alpha mn} & \mat{R}
\end{bmatrix} \in \ff{2}^{\alpha m n \times m n},~\mat{R} \in \ff{2}^{\alpha m n \times (1-\alpha)m n}$, where $\rk{(\mat{R})} = n \rk{(\mat{P})}  = n\min{(\alpha m ,(1-\alpha)m)}$. Since in our setting $\alpha > \frac{7}{8} > \frac{1}{2}$, the linear map $\ff{2}^{\alpha m n} \rightarrow \ff{2}^{(1-\alpha) m n}$ defined by the matrix $\mat{R}$ is surjective. Now, let $Z_{\ff{2}}(\mathcal{P})$ be the set of solutions to $\mathcal{P}$ over $\ff{2}$. We can express it as the Cartesian product $V \times W$, where $V \subset \ff{2}^{\alpha m n}$ and $W \subset \ff{2}^{(1-\alpha)mn}$ are the solution sets of $\lbrace p_1,\dots,p_{\alpha m} \rbrace$ and $\lbrace p_{\alpha m+1},\dots,p_m \rbrace$, respectively. Since the solutions of $\mathcal{F}$ correspond to the dual Reed-Solomon codewords that satisfy the equations of $\mathcal{P}$, we obtain the final solution set
\begin{equation*}
    Z_{\ff{2}}(\mathcal{F}) = \left\lbrace (\mbf{a},\mbf{a}\mat{R}),~\mbf{a}\in V,~\mbf{a}\mbf{R} \in W\right\rbrace.
\end{equation*}
In particular, we see that the right factor of each element of $Z_{\ff{2}}(\mathcal{F})$ belongs to $V\mbf{R} \cap W$. This makes it important to grasp this intersection, noting that $\mbf{R} \in \ff{2}^{\alpha mn \times (1-\alpha)mn}$ depends on the Reed-Solomon code and the extension field basis, while $V \times W$ is the set of solutions of $m$ Boolean degree-$d$ polynomials sampled independently from this code. As the linear map $\ff{2}^{\alpha m n} \rightarrow \ff{2}^{(1-\alpha) m n}$ defined by $\mat{R}$ is surjective and as $\abs{V} = 2^{\alpha m(n-1)}$ remains significantly larger than $\abs{\ff{2}^{(1-\alpha)mn}} = 2^{(1-\alpha)m n}$, we adopt the following assumption, which can also be verified experimentally. Under this assumption, we have that $V\mbf{R} \cap W = W$.
\begin{assumption}\label{ass:VR}
Using the notation above, we assume that the linear map defined by $\mat{R}$, when restricted to $V$, remains surjective.
\end{assumption}

\begin{theorem}\label{theorem:best_combinatorial}
    Under Assumption~\ref{ass:VR}, there exists a classical algorithm that solves the system $\mathcal{F}$ in time $\widetilde{\mathcal{O}}(2^{(1-\alpha)m}) = \widetilde{\mathcal{O}}(2^{(1-\alpha)n^2})$.
\end{theorem}
\begin{proof}
   First, we can neglect the complexity of finding all solutions to $\mathcal{P}$ since it is lower than the cost claimed in the theorem. Indeed, we can simply solve the $m$ degree-$d$ polynomials of $\mathcal{P}$ separately, as each involves $n$ distinct variables. Note again that the degree $d$ is treated as a constant.
   
   Then, using the notation above, the $\alpha m n$ leftmost components of an arbitrary element in $Z_{\ff{2}}(\mathcal{P})$ correspond to a vector $\mat{a} \in V$. This element is an element of $Z_{\ff{2}}(\mathcal{F})$ if and only if the vector $\mat{a}\mat{R}$ belongs to $W$. This happens with probability $$\frac{\abs{V\mat{R} \cap W}}{ \abs{V\mat{R}}} = \frac{\abs{W}}{\abs{\ff{2}^{(1-\alpha)mn}}} = \frac{2^{(1-\alpha)m(n-1)}}{2^{(1-\alpha)mn}} = 2^{(\alpha-1)m},$$
   where the first equality is under Assumption \ref{ass:VR}. The cost follows by taking the inverse of this probability.
\end{proof}
Note that another naive algorithm is to sample elements of $C$ and test if they satisfy the equations of $\mathcal{P}$, but with worse complexity $$\widetilde{\mathcal{O}}{\left(\frac{\abs{C}}{\abs{Z_{\ff{2}}(\mathcal{F})}}\right)} = \widetilde{\mathcal{O}}\left(\frac{2^{\alpha m n}}{2^{\alpha m n-m}}\right) = \widetilde{\mathcal{O}}(2^{m}) = \widetilde{\mathcal{O}}(2^{n^2}).$$ 
 Finally, even if the sets $V$ and $W$ can be assumed to be known, we have not found any clever collision-based methods for finding elements in $V\mat{R} \cap W$. In fact, under Assumption \ref{ass:VR}, such methods do not exist.

\subsection{Gröbner basis algorithms}\label{sec:generic}

Gröbner basis methods can often be more efficient than other types of solvers for certain structured systems, both empirically and in some provable cases. Before applying them, we restrict ourselves to the setting typically considered in cryptanalysis, namely, overdetermined systems. For that purpose, we fix variables in $\mathcal{F}$ to reduce it to a smaller overdetermined degree-$d$ system $\mathcal{F}_{\text{spec}}$ with $\mathcal{O}(n^2)$ equations and variables, as was done at the beginning of Section \ref{sec:combi}. The goal is to determine whether Gröbner basis methods can offer a significant improvement over exhaustive search and enumerative methods for such an $\mathcal{F}_{\text{spec}}$.

It is known that solving a generic degree-$d$ system with the same number of equations and variables as $\mathcal{F}_{\text{spec}}$ would require time exponential in $n^2$, with precise estimates for the associated complexity exponent available (see, for instance,~\cite {bardet_these,BFS15}). However, our system $\mathcal{F}_{\text{spec}}$ is highly structured. As we have just said, this structure may be what makes Gröbner basis methods significantly more efficient, while also making their theoretical analysis more difficult. In our context, the structure of $\mathcal{F}_{\text{spec}}$ depends on both the structure of the original system $\mathcal{F}$ and the specific way in which the specialization is carried out.

It is extremely unclear how to approximate the complexity of solving $\mathcal{F}_{\text{spec}}$ by that of a generic degree-$d$ system, or by that of any other structured system whose solving complexity has been tackled in the literature. We provide some theoretical and experimental evidence in Appendix~\ref{app:GB} that points to the exponential run time required by some common Gröbner basis algorithms. We report the timings of experiments considering one specialization strategy on $\mathcal{F}$, which seems to be the most relevant, examining the performance of Gröbner basis algorithms on the resulting specialization $\mathcal{F}_{\text{spec}}$. We refer readers to Appendix~\ref{app:GB} for details.




\section*{Acknowledgements}
We sincerely thank Swastik Kopparty  for his gracious assistance in pointing us to~\cite{lidl1997finite} and the normal form for quadratic polynomials contained therein, and for enlightening conversations over the years! 


\pagebreak
\bibliographystyle{alpha}
\bibliography{Bibliography/abbrev0, Bibliography/crypto, Bibliography/custom}

\appendix
\section{Details on Gröbner Basis Algorithms}\label{app:GB}

This appendix gives more details on solving strategies that involve specializing variables in the system $\mathcal{F}$, followed by the application of Gröbner basis techniques to the resulting mildly overdetermined system $\mathcal{F}_{\text{spec}}$. In Section \ref{sec:grobner}, we describe these techniques before applying them to such an $\mathcal{F}_{\text{spec}}$. Their cost depends on several factors, such as the number of equations, the number of variables, and their degrees, but it is ultimately governed by the system's overall structure, whose influence is also the most difficult to characterize. In Section~\ref{sec:F_to_Fspec}, we detail the structure of $\mathcal{F}_{\text{spec}}$. As established in Section~\ref{sec:quads_easy}, we have already ruled out the existence of clever specializations that would yield equations of degree lower than $d$, so that $\mathcal{F}_{\text{spec}}$ only contains degree-$d$ equations. However, the way specialization is carried out still influences the precise number of equations and variables beyond the $\mathcal{O}(n^2)$ baseline, as well as the overall structure of the system. In Section~\ref{sec:magma}, we present the results of our experiments for one particular specialization strategy, which seems to yield the easiest system to solve.

\subsection{Solving overdetermined systems via Gröbner bases}\label{sec:grobner}

Broadly speaking, by applying Gröbner basis techniques, we mean either applying directly standard Gröbner basis algorithms commonly employed in cryptanalysis~\cite{f4,f5} or combining them with additional specializations that aim to make the system $\mathcal{F}_{\text{spec}}$ more overdetermined, thereby facilitating their application. Algorithms from~\cite{bettale2009,xbred,bdt} fall into the second category, which we refer to as ``hybrid methods''. Since we start from a system $\mathcal{F}_{\text{spec}}$ having an equal number of equations and variables, the probability that a random specialization keeps this system consistent is exponentially small in the number of specialized variables. Therefore, the process must be repeated approximately as many times as the inverse of that probability, with consistency checked in each iteration by applying the Gröbner basis algorithm. 

Both types of methods have exponential complexity in the MQ regime that is generally considered in cryptanalysis, and solving the same number of generic degree-$d$ equations in the same number of variables for $d \geq 3$ is even more costly than in the quadratic case~\cite{bardet_these,BFS15}. Note also that an asymptotic comparison between these two methods has only been sketched for generic MQ systems (see \cite{bettale2009}). While the work in~\cite{bettale2009} can, in principle, be extended to higher-degree generic systems, such a comparison has not yet been performed for more structured systems.

Our experiments in Section \ref{sec:magma} include the simplest hybrid method of \cite{bettale2009} by fixing randomly $f \geq 1$ variables in $\mathcal{F}_{\text{spec}}$. Our goal there is to explore whether variable fixing can lead to empirical improvements rather than to provide a full asymptotic comparison. In fact, analyzing Gröbner basis algorithms applied directly to $\mathcal{F}_{\text{spec}}$, without fixing any variables, is already quite challenging (this first task can be seen both as a particular case and as a prerequisite for such a comparison).

\subsection{Reduction to an overdetermined system $\mathcal{F}_{\text{spec}}$}\label{sec:F_to_Fspec}

In the following, we set $k = \alpha m$ for the dimension of the dual Reed-Solomon code over $\ff{2^n}$, and we reuse some notation of Section \ref{sec:combi}. Specifically, let
\begin{equation}\label{eq:barG}
    \overline{\mat{G}} = \begin{bmatrix}
\mat{I}_{kn} & \mat{R}
\end{bmatrix} \in \ff{2}^{k n \times m n},~\mat{R} \in \ff{2}^{k n \times (m-k)n}
\end{equation}
be a generator matrix for the binary image of this code, once again obtained by choosing an appropriate extension field basis. In Section \ref{sec:combi}, we have seen that the matrix $\mat{R}$ is of full rank in this case. Let us also recall that the system $\mathcal{F}$ was obtained from the initial degree-$d$ equations of $\mathcal{P}$ by eliminating variables using the parity check equations. This process amounts to introducing a vector of $k n$ new variables $\mat{y} = (\mat{y}_1,\dots,\mat{y}_{k})$ such that $\mat{y}\overline{\mat{G}} = \mat{x}$, where $\mat{x}$ represents the original variables. The system $\mathcal{F}$ is then reinterpreted as a system in these new variables $\mat{y}$. More precisely:
\begin{system}[Underdetermined system $\mathcal{F}$]\label{sys:under_F}
Let $\mathcal{P} = \lbrace p_1,\dots,p_m \rbrace$ denote the initial system, where $p_i \in \ff{2}[\mat{x}_i]$ for $i \in [m]$, let $\overline{\mat{G}} = \begin{bmatrix}
\mat{I}_{kn} & \mat{R}
\end{bmatrix} \in \ff{2}^{kn \times mn}$ denote the generator matrix of Equation \eqref{eq:barG}, and let $\mat{R}_{j}$ denote the $j$-th block of size $k n \times n$ in $\mat{R}$ for $j \in [m-k]$. The equations of $\mathcal{F} \subset \ff{2}[\mat{y}_1,\dots,\mat{y}_{k}]$ can be defined by:
\begin{itemize}
    \item for $i \in [k]$, $$f_i(\mat{y}_i) = p_i(\mat{y}_i);$$
    \item for $i \in [k+1..m]$, $$f_i(\mat{y}) = p_i(\mat{y}\mat{R}_{i-k}).$$
\end{itemize}
In other words, $k$ equations of $\mathcal{F}$ remain unchanged from those in $\mathcal{P}$, while the remaining $m-k$ equations involve all $y_{i,j}$ variables mixed in a way that depends on the dual Reed-Solomon code.
\end{system}

We now describe specialization strategies applied to $\mathcal{F}$ that yield an overdetermined system $\mathcal{F}_{\text{spec}}$. The main point is that we are free to choose how to distribute the specializations across the $\mat{y}_i$ blocks.

\begin{system}[Specialization of System \ref{sys:under_F}]\label{sys:well_F} Let $J \subset [k n]$ be the set of specialized positions in $\mat{y}$ and let $\mat{t} \in \ff{2}^{\abs{J}}$ be the vector that contains the specialized values. In other words, each initial block $\mbf{y}_i$ is replaced with a block whose components can be reordered as $(\mat{z}_i,\mat{t}_i)$, where $\mat{t}_i = \mat{t}_{j \in J \cap [n(i-1)+1..ni]}$ contains binary elements and where $\mat{z}_i = \mat{z}_{j \in \bar{J} \cap [n(i-1)+1..ni]}$ still contains variables. Note that both $\mat{z}_i$ and $\mat{t}_i$ can be empty if all and no positions are fixed within this block, respectively. We consider $\mat{z} = (\mat{z}_1,\dots,\mat{z}_{k'})$ the set of non-empty blocks, where $k' = \abs{\lbrace i \in [k],~\abs{J \cap [n(i-1)+1..ni]} < n \rbrace}$. The specialization of $\mathcal{F}$ at the positions $J$ to the values described by $\mat{t} \in \ff{2}^{\abs{J}}$ is then given by:
\begin{itemize}
    \item for $i \in [k']$,
\begin{equation}\label{eq:one_block}
	h_i(\mbf{z}_i) = p_i(\mbf{y}_i)|_{\mat{y}_{j \in J \cap [n(i-1)+1..ni]} = \mat{t}_{i}},
\end{equation}
\item for $i \in [k + 1..m]$,
\begin{equation}\label{eq:all_blocks}
	h_i(\mbf{z}) = p_i(\mat{z}\mat{S}_{i-k} + \underbrace{\mat{t}\mat{T}_{i-k}}_{{\in \ff{2}}}),
\end{equation}
\end{itemize}
where $\mat{S}_{i} \in \ff{2}^{(kn - \abs{J}) \times n}$ is the set of rows of the matrix $\mat{R}_{i}$ defined in System \ref{sys:under_F} with indexes in $\bar{J}$ and where $\mat{T}_{i} \in \ff{2}^{\abs{J} \times n}$ is the set of remaining rows.
\end{system}

\begin{remark}
    We note that our specialized systems share similarities with hard-to-solve algebraic systems considered in recent cryptanalyses~\cite{regular_sd,regular_mq}. Indeed, they contain constraints involving specific blocks of variables $\mat{z}_i$, as well as independent constraints involving all variables $\mat{z}$, similar to the structure observed in those works. However, there are significant differences. First, the degree of our equations exceeds those in~\cite{regular_sd,regular_mq}, where all equations are of degree 1 or 2. Moreover, the sizes of the variable blocks can vary in our case, unlike in~\cite{regular_sd,regular_mq}. Most notably, the variables $\mat{z}$ are compressed into length-$n$ vectors via the $\mat{S}_i$ matrices in Equation~\eqref{eq:all_blocks}, which prevents us from directly treating the constraints involving all variables as generic degree-$d$ constraints. This contrasts with the generic degree-1 (resp. degree-2) constraints considered in~\cite{regular_sd} (resp. \cite{regular_mq}).
\end{remark}
Let $n_i \in [n]$ be the size of $\mat{z}_i$ from System \ref{sys:well_F} for $i \in [k']$. Specializations that yield a well-defined system correspond to the condition
\begin{equation*}
    \sum_{i = 1}^{k'} n_{i} = m - k + k'.
\end{equation*}
Our $\mathcal{F}_{\text{spec}}$ will be a system of the type described in System \ref{sys:well_F} that satisfies this criterion. Regardless of the choice of $J$, this implies that the number of equations and variables is in $[m-k..m]$, and thus is in $\Theta(n^2)$.

We will mostly discuss $J$ rather than the choice of the corresponding set of specialized values $\mat{t} \in \ff{2}^{\abs{J}}$. Different fixing strategies, along with partial or full knowledge of the solutions to $\mathcal{P}$, can result in different success probabilities. However, as our main goal is to determine if polynomial-time algorithms exist for solving $\mathcal{F}_{\text{spec}}$, we will not explore this aspect in detail. In particular, we expect that different specializations over the same $J$ will require roughly the same amount of time to solve. Among choices of $J$ in which the final blocks $\mat{z}_i$ all have the same length, one natural configuration is to fix variables evenly across the $k$ blocks of $\mat{y}$, resulting in each block having $n_i = m/k = 1/\alpha$ unfixed variables for $i \in [k]$. Concretely, this implies that, on average, each $\mat{z}_i$ contains one or two unfixed variables. Another configuration corresponds to fixing as many blocks as possible completely, leaving the remaining ones unchanged. In this case, the number of remaining (fully unfixed) blocks is given by $k' = \frac{m-k}{n-1}$. Naturally, there are trade-offs, including approaches whose final blocks have different lengths. While some particular choices of $J$ may offer advantages in terms of solving complexity, the way the solving algorithms of Section \ref{sec:grobner} operate suggests that the asymptotic behavior should remain roughly the same. In other words, we expect that a polynomial-time algorithm will either exist for all such specialized systems or none of them.

The experiments reported below focus on the second configuration with $\frac{(1-\alpha)m}{n-1} = \frac{m-k}{n-1}$ blocks, each of size $n$. Recalling that $n\frac{m-k}{n-1} \approx (1-\alpha)n^2$, this specialized system has the size outlined at the beginning of Section~\ref{sec:combi}. In particular, it is smaller by a constant factor, specifically $1 - \alpha \leq 1/8$, compared to the system arising from the first configuration in which the variables are evenly distributed into $k$ blocks $\mat{z}_i$, each of size about $1/\alpha$. This motivated our decision to focus our experiments on the second configuration.

\subsection{Gröbner basis experiments on $\mathcal{F}_{\text{spec}}$}\label{sec:magma}

For $d=3$ and for several values of $n$, $m=n^2$, and $\alpha \geq 7/8$, we performed experiments on the system $\mathcal{F}_{\text{spec}}$ obtained via the specialization approach just described above with $\frac{(1-\alpha)m}{n-1} = \frac{m-k}{n-1}$ blocks, each of size $n$. For the inner Gröbner basis algorithm, we used the version of Faugère’s F4 algorithm~\cite{f4} implemented in Magma~\cite{magma}. The value of $f$ is the number of extra variables fixed in $\mathcal{F}_{\text{spec}}$: specifically, $f=0$ corresponds to plain Gröbner basis computation, while $f>0$ indicates hybrid Gröbner basis methods. Each table cell reports the number of equations $n_{\text{eqs}}$ and variables $n_{\text{vars}}$ in $\mathcal{F}_{\text{spec}}$, written as $(n_{\text{eqs}}, n_{\text{vars}})$ and the time to solve the system in seconds. We made sure to perform enough tests for one single parameter set, especially when $f > 0$. Indeed, hybrid Gröbner methods proceed by solving several inconsistent systems before finding a consistent one, and thus variation in running time increases as $f$ grows. Cells with ``memory'' indicate an aborted computation due to excessive memory consumption.

\begin{table}[H]
\begin{center}
\begin{tabular}{ |p{0.5cm}||p{3.1cm}|p{3.1cm}|p{3.1cm}|p{3.1cm}| }
 \hline
 $n$ & $f=0$ & $f=1$ & $f=2$ & $f=3$ \\
 \hline
 7 & (9,9)\quad 0.131 &&& \\
 8 & (10,10)\quad 0.299 &&& \\
 9 & (13,13)\quad 2.460 & (13,12)\quad 3.526 & (13,11)\quad 5.984 & (13,10)\quad 9.020 \\
 10 & (15,15)\quad 9.253 & (15,14)\quad 11.797 & (15,13)\quad 17.925 & (15,12)\quad 34.235 \\
 11 & (18,18)\quad 239.157 & (18,17)\quad 66.645 & (18,16)\quad 116.734 & (18,15)\quad 135.665 \\
 12 & (20,20)\quad 1813.301 & (20,19)\quad 1046.992 & (20,18)\quad 563.290 & (20,17)\quad 580.349 \\
 13 & (24,24)\quad memory &&& (24,21)\quad 9023.630 \\
 \hline
\end{tabular}
\caption{Experiments for $\alpha = 0.875$.}
\label{tab:full_blocks_small_a}
\begin{tabular}{ |p{0.5cm}||p{3.1cm}|p{3.1cm}|p{3.1cm}|p{3.1cm}| }
 \hline
 $n$ & $f=0$ & $f=1$ & $f=2$ & $f=3$ \\
 \hline
 9 & (11,11)\quad 0.918 & (11,10)\quad 1.430 & (11,9)\quad 2.668 & (11,8)\quad 3.936 \\
 
 10 & (12,12)\quad 2.057 & (12,11)\quad 3.994 & (12,10)\quad 7.095 & (12,9)\quad 9.345 \\
 
 11 & (15,15)\quad 11.229 & (15,14)\quad 15.710 & (15,13)\quad 33.378 & (15,12)\quad 54.194 \\
 
 12 & (17,17)\quad 69.832 & (17,16)\quad 48.698 & (17,15)\quad 87.304 & (17,14)\quad 134.562 \\
 
 13 & (19,19)\quad 528.655 & (19,18)\quad 365.466 & (19,17)\quad 485.759 & (19,16)\quad 500.332 \\

 14 & (22,22)\quad memory & (22,21)\quad 7887.640 & (22,20)\quad 3939.470 & (22,19)\quad 5415.453 \\
 \hline
\end{tabular}
\caption{Experiments for $\alpha = 0.9$.}
\label{tab:full_blocks_mid_a}
\begin{tabular}{ |p{0.5cm}||p{3.1cm}|p{3.1cm}| }
 \hline
 $n$ & $f=0$ & $f=1$  \\
 \hline
 9 & (8,8)\quad 0.167 & (8,7)\quad 0.271 \\
 10 & (9,9)\quad 0.730 & (9,8)\quad 0.583 \\
 11 & (11,11)\quad 1.506 & (11,10)\quad 1.479 \\
 12 & (12,12)\quad 3.620 & (12,11)\quad 4.824 \\
 13 & (15,15)\quad 36.173 & (15,14)\quad 27.588 \\
 14 & (17,17)\quad 102.314 & (17,16)\quad 103.471 \\
 15 & (19,19)\quad 461.174 & (19,18)\quad 365.748 \\
 16 & (22,22)\quad 4229.740 & (22,21)\quad 3565.840 \\
 \hline
\end{tabular}
\caption{Experiments for $\alpha = 0.925$.}
\label{tab:full_blocks_mid_mid_a}
\end{center}
\end{table}
\begin{table}[H]
\begin{center}
\begin{tabular}{ |p{0.5cm}||p{3.1cm}|p{3.1cm}|p{3.1cm}|p{3.1cm}| }
 \hline
 $n$ & $f=0$ & $f=1$ & $f=2$ & $f=3$ \\
 \hline
9 & (6,6) \quad 0.069 & (6,5) \quad 0.088 & (6,4) \quad 0.181 & (6,3) \quad 0.283 \\
10 & (6,6) \quad 0.083 & (6,5) \quad 0.107 & (6,4) \quad 0.202 & (6,3) \quad 0.488 \\
11 & (8,8) \quad 0.403 & (8,7) \quad 0.892 & (8,6) \quad 0.960 & (8,5) \quad 2.464 \\
12 & (9,9) \quad 1.177 & (9,8) \quad 2.007 & (9,7) \quad 3.808 & (9,6) \quad 7.678 \\
13 & (10,10) \quad 3.386 & (10,9) \quad 7.014 & (10,8) \quad 14.670 & (10,7) \quad 19.684 \\
14 & (11,11) \quad 7.536 & (11,10) \quad 11.720 & (11,9) \quad 28.467 & (11,8) \quad 52.197 \\
15 & (13,13) \quad 25.164 & (13,12) \quad 60.169 & & \\
16 & (14,14) \quad 47.116 & (14,13) \quad 94.921 & & \\
17 & (16,16) \quad 101.875 & (16,15) \quad 174.440 & & \\
18 & (18,18) \quad 359.204 & (18,17) \quad 417.592 & & \\
19 & (21,21) \quad 1981.867 & (21,20) \quad 3162.036 & & \\
20 & (22,22) \quad memory & (22,21) \quad 6142.053 & & \\
 \hline
\end{tabular}
\caption{Experiments for $\alpha = 0.95$.}
\label{tab:full_blocks_big_a}
\end{center}
\end{table}
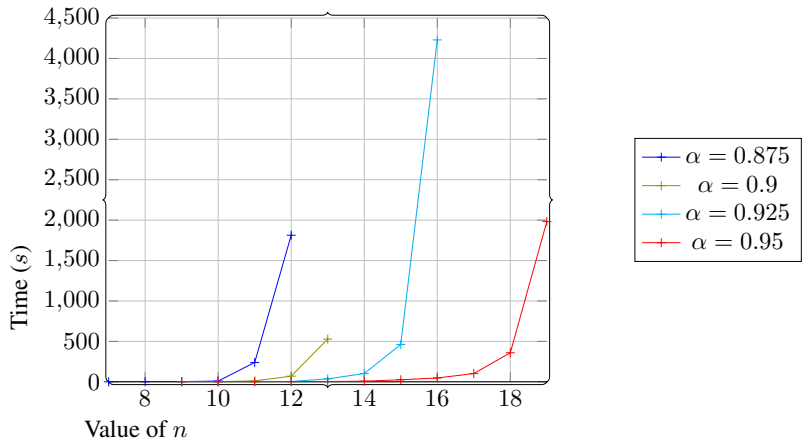
\begin{figure}[h]
\centering

\pgfplotstableread[row sep=\\]{
X Y \\ 7 0.131 \\ 8 0.299 \\ 9 2.460 \\ 10 9.253 \\ 11 239.157 \\  12 1813.301 \\  
}{\dataExpPP}

\pgfplotstableread[row sep=\\]{
X Y \\ 9 0.918 \\ 10 2.057 \\ 11 11.229 \\ 12 69.832 \\ 13 528.655 \\ 
}{\dataExpP}

\pgfplotstableread[row sep=\\]{
X Y \\ 9 0.167 \\ 10 0.730 \\ 11 1.506 \\ 12 3.620 \\ 13 36.173 \\ 14 102.314 \\ 15 461.174 \\ 16 4229.740 \\ 
}{\dataExpF}

\pgfplotstableread[row sep=\\]{
X Y  \\ 9 0.069 \\  10 0.083 \\  11 0.403 \\  12 1.177 \\ 13 3.386 \\ 14 7.536 \\ 15 25.164 \\ 16 47.116 \\ 17 101.875 \\ 18 359.204 \\ 19 1981.867 \\
}{\dataExpFF}

\begin{tikzpicture}[scale=0.85]
\pgfplotsset{every axis legend/.append style={at={(1,0.2)},anchor=east}}
\def\nmin{7}\def\nmax{19}
\def\Cmin{0}\def\Cmax{4500}
\begin{axis}[ grid=major, try min ticks=9,
xmin=\nmin, xmax=\nmax, ymin=\Cmin, ymax=\Cmax,
xlabel style={at={(0.2,-0.13)},anchor=east},
ylabel style={at={(-0.2,0.4)},anchor=east},
xlabel={Value of \(n\)},
ylabel={Time ($s$)},
legend style={ at={(1.2,0.5)}, anchor=west },
title={}
]

\addplot[blue,mark=+] table {\dataExpPP};
\addlegendentry{$\alpha = 0.875$}
\addplot[olive,mark=+] table {\dataExpP};
\addlegendentry{$\alpha = 0.9$}

\addplot[cyan,mark=+] table {\dataExpF};
\addlegendentry{$\alpha = 0.925$}

\addplot[red,mark=+] table {\dataExpFF};
\addlegendentry{$\alpha = 0.95$}

\end{axis}

\end{tikzpicture}
\caption{\label{fig:r=345exp}Time complexity of plain Gröbner basis approach ($f=0$).}\label{fig:comp}
\end{figure}

\paragraph{Interpretation.} Due to the small value of $n$, we do not attempt to infer an exact time complexity from these results. Nevertheless, our experiments support exponential growth in $(1-\alpha)n^2$ for the plain Gröbner basis approach. Furthermore, the observation that hybrid Gröbner basis approaches for mild $f>0$ are sometimes slower and sometimes faster than the plain Gröbner basis approach suggests that, overall, hybrid Gröbner basis techniques do not necessarily have a consistent advantage over combinatorial algorithms studied in Section \ref{sec:combi}, where $\mathcal{F}_{\text{spec}}$ was solved via pure enumerative methods, for \emph{all} parameter sets $n,~\alpha$, and even $d$.

\section{Leveraging the Code Structure in Classical Cryptanalysis}\label{app:exploiting_C}

We have tried to exploit the efficient decoding procedure or, at least, the structural properties of the Reed-Solomon code for classical cryptanalysis. Before describing some unsuccessful attempts, note that, according to our analysis, the complexity of the best solving algorithms studied in Sections \ref{sec:combi} and \ref{sec:generic} depends only on the code rate $\alpha$.

Using Lemma \ref{lem:dualrs}, our dual Reed-Solomon code is a particular Generalized Reed-Solomon code. Each of its codewords $\mat{c} \in C$ can thus be written as 
\begin{equation*}
	\mat{c} = (v'_1g(\gamma_1),\dots,v'_m g(\gamma_m)) \in \ff{2^n}^m,
\end{equation*}
where $(v'_1,\dots,v'_m)$ and $(\gamma_1,\dots,\gamma_m)$ are fixed and where $g$ is a univariate polynomial of degree $<k$. Using a fixed $\ff{2}$-isomorphism between and $\ff{2}^n$ and $\ff{2^n}$, it is also standard to associate a set of $n$ quadratic Boolean polynomials $\lbrace f_1,\dots,f_n \rbrace$ in $n$ variables $x_1,\dots,x_n$ with a univariate polynomial $F$ over $\ff{2^n}$ in a new variable $X$, such that this polynomial is quadratic in terms of $(X,\dots,X^{2^{n-1}})$. The same holds if the $f_i$'s have degree $d \geq 3$, and this time the polynomial $F$ has degree $d$ in $(X,\dots,X^{2^{n-1}})$. Doing this on $\lbrace p_i,0\dots,0\rbrace$ for $i \in [m]$ from our construction yields a polynomial system $\lbrace P_1,\dots,P_m\rbrace \subset \ff{2^n}[X_1] \times \dots \times \ff{2^n}[X_m]$, whose solutions correspond to that of $\mathcal{P}$ via the fixed isomorphism. Our problem can thus be restated as the one of finding a univariate polynomial $g$ of degree $<k$ such that $P_i(v'_ig(\gamma_i)) = 0$ for $i \in [m]$. By introducing $k$ formal unknowns corresponding to the coefficients of $g$, this gives a polynomial system of $m$ equations and $k$ variables over $\ff{2^n}$. A system of the same type can in fact be derived from an arbitrary $\ff{2^n}$-linear code with generator matrix $\mat{G} \in \ff{2^n}^{k \times n}$ by introducing variables $y_i \in \ff{2^n}$ for a codeword $(y_1,\dots,y_k)\mat{G}$. Therefore, such an approach should not bring direct improvements.

More generally, we discuss the feasibility of using the efficient decoding procedure of any type of structured code, beyond just Reed-Solomon. First, it is not obvious to us how to formulate a relevant decoding instance -- target point, metric -- that yields a codeword whose components satisfy the equations of $\mathcal{P}$. However, we can still hope to use the structure of $C$ to efficiently find such a codeword, perhaps more directly, as we have just tried to achieve in the case of a Generalized Reed-Solomon code. When the code is unstructured, decoding algorithms and algorithms to find low-weight codewords are fairly similar (the same happens for random lattices). Therefore, a first optimistic scenario for cryptanalysis might be that the polynomial constraints -- which are independent of $C$ -- model a sort of low-weight feature or more generally a rather simple-to-state property. As far as we know, this does not seem to be the case. To see this, let us assume for now that $C$ still comes from an $\ff{2^n}$-linear code $\subset \ff{2^n}^m$. In this case, the set of solutions to $\lbrace P_1,\dots,P_m\rbrace$ is the product $Z = Z_1 \times \dots \times Z_m$, where $Z_i \subset \ff{2^n}$ are the roots of the univariate polynomial $P_i$ and the $Z_i$'s are independent of each other. While one could try to exploit the non-random character of each $Z_i \subset \ff{2^n}$, the coordinate-wise independence does not seem advantageous from a solving perspective. A second optimistic scenario is that there exist code families tailored to such a set $Z$, but obviously, the code chosen to instantiate the construction has no reason to belong to such a family. Overall, all these reasons make us think that an efficient classical algorithm that can exploit the structure of $C$ should be a significant contribution to cryptanalysis.

\end{document}